\documentclass[preprint]{ptephy_v1}

\usepackage{boites}
\usepackage{amsmath}
\usepackage{amsthm}
\usepackage{mathtools}
\usepackage{mathrsfs}
\usepackage{color}
\usepackage{braket}
\usepackage{url}

\newcommand{\del}{\partial}
\newcommand{\ua}{\uparrow}
\newcommand{\da}{\downarrow}
\newcommand{\basis}{\boldsymbol{r}\, \sigma}
\newcommand{\basissp}{\boldsymbol{r}\, \sigma'}

\newcommand{\e}{\mathrm{e}}
\newcommand{\vap}{\varphi_{2,i}^{q+}}
\newcommand{\vam}{\varphi_{2,i}^{q-}}
\newcommand\ii{\mathrm{i}}
\newcommand{\diag}{\operatorname{diag}}
\newcommand{\pf}{\operatorname{pf}}
\newcommand{\sgn}{\operatorname{sgn}}
\renewcommand{\Re}{\operatorname{Re}}
\renewcommand{\Im}{\operatorname{Im}}

\preprintnumber{YITP-20-67, KUNS-2816}

\begin{document}

\title{Hartree--Fock--Bogoliubov theory for odd-mass nuclei with a time-odd constraint and 
application to deformed halo nuclei}

\author{Haruki Kasuya\thanks{Email: haruki.kasuya@yukawa.kyoto-u.ac.jp}}
\affil{Yukawa Institute for Theoretical Physics, Kyoto University, Kyoto 606-8502, Japan}
\author{Kenichi Yoshida\thanks{Email: kyoshida@ruby.scphys.kyoto-u.ac.jp}}
\affil[2]{Department of Physics,  Kyoto University, Kyoto 606-8502, Japan}

\begin{abstract}
We 
show that 
the lowest-energy solution of the Hartree--Fock--Bogoliubov (HFB) equation has the even particle-number parity 
as long as the time-reversal symmetry is conserved in the HFB Hamiltonian without null eigenvalues. 
Based on this finding, we give a rigorous foundation of 
a method for solving the 
HFB equation to describe the ground state of odd-mass nuclei by 
employing a time-reversal anti-symmetric constraint operator to the Hamiltonian,  where 
one obtains directly the ground state  
as a self-consistent solution of the cranked-HFB-type equation. 
Numerical analysis is done for the neutron-rich Mg isotopes with a reasonable choice for the operator, 
and it is demonstrated that
the anomalous increase in the matter radius of $^{37}$Mg is well described 
when the last neutron occupies a low angular-momentum orbital 
in the framework of the nuclear energy-density-functional 
method, revealing the deformed halo structure. 

\end{abstract}

\subjectindex{D10, D11, D13}

\maketitle

\section{Introduction}\label{sec1}
Odd-mass nuclei, composed of an odd number of 
nucleons, 
unveil the unique features that one cannot observe in even--even nuclei. 
Even--even nuclei, for instance, have spin $J^\pi=0^+$ in the ground state, 
where nucleons are paired off due to the correlation, 
while odd-mass nuclei have nonzero ground-state spin, 
where the last nucleon does not take part in the pair correlation and 
is responsible for the total spin. 
The spin gives us information of single-particle orbitals near the Fermi level. 
With increase in the structure information of unstable nuclei 
thanks to the recent advancement in radioactive-isotope beam technology~\cite{nak17}, 
more and more exotic features in odd-mass nuclei have been showing up: 
Highlights in the latest discoveries include 
the shape staggering in $^{181\textrm{--}185}$Hg~\cite{mar18}, and 
the deformed halo structures of $^{31}$Ne~\cite{nak09,tak12} and $^{37}$Mg~\cite{tak14,kob14}.  
A great theoretical challenge under these circumstances 
is to describe odd-mass nuclei in a wide mass region of nuclear chart, 
where the pair correlation, the shape deformation, and the weak binding of nucleons are all 
considered in a unified manner. 

Nuclear density-functional theory (DFT) or the self-consistent mean-field model 
has been extensively employed for describing the systematic features of not only the ground 
but excited states~\cite{ben03,nak16}. 
The nuclear landscape has been investigated in the framework of both non-relativistic and relativistic 
energy-density functional (EDF) methods~\cite{erl12, afn13}. 
The Hartree--Fock--Bogoliubov (HFB) theory or the Kohn--Sham--Bogoliubov--de-Gennes scheme in DFT is 
capable of providing us a unified description of the ground-state properties for not only even--even nuclei but odd-mass nuclei 
taking the superfluidity and shape deformation into account~\cite{rin80}. 
In spite of the successful application of DFT, 
the calculations have been mostly restricted to even--even nuclei, 
and odd-mass nuclei remain largely unexplored~\cite{sch10}. 
This may be partly because the primal interest has been put on 
determining the drip lines~\cite{sto03}. 
With the recent advent of computational resources sufficient to perform global calculation in the framework of DFT, 
systematic odd-even alternations in atomic nuclei such as the odd-even staggering of the binding energies have attracted renewed interest~\cite{ber09}.

There seem many obstacles to tackling the systematic investigation of odd-mass nuclei in DFT, 
and some of them are the followings:  
(i) Coexistence of multiple levels in low energy;  
to excite even-even nuclei we have to break at least one pair of nucleons whose binding energy is of the order of 1--2 MeV, whereas for odd-mass nuclei an excitation can be achieved by putting up the last nucleon by a few hundred keV.
As a result, 
one-quasiparticle and phonon states appear in the low-excitation energy and they can admix with one another~\cite{kis63}, 
and thus the HFB describes better the ground state of even-even nuclei.
The high precision calculation, at the same time with high accuracy, 
is also required for resolving the near degeneracy of several levels 
and identifying the ground state for odd-mass nuclei. 
This is not restricted to DFT but a challenge for any theoretical models.
(ii) Non-vanishing time-odd densities; 
an EDF is a time-even scalar constructed from various densities, 
and includes the densities and currents that are odd with respect to time reversal to preserve the Galilean or Lorentz invariance and 
to describe properly e.g. the spin-dependent observables. 
While the time-odd densities and the related time-odd fields automatically vanish for the ground state of 
even--even nuclei, 
they are nonzero in odd-mass nuclei where the time-reversal symmetry 
is intrinsically broken~\cite{ben03}. 
Allowing the breaking of time-reversal symmetry increases the computational cost~\cite{per08}. 
Furthermore, EDFs commonly used in the practical calculations are phenomenologically constructed 
by using the properties of time-even states only. 
Thus, in the non-relativistic case, the coupling constants of time-odd fields are highly uncertain.  
In the relativistic case, on the other hand, the coupling constants of time-odd fields are defined from those of time-even fields through the Lorentz invariance, so there are no such uncertainties~\cite{hof88, afa00, afa10, afa11} 
(iii) Complexity of the blocking procedure; 
one cannot usually obtain the ground state of odd-mass nuclei 
as the lowest-energy solution 
and 
needs an additional procedure to excite one quasiparticle 
on top of the ground state of even--even nuclei~\cite{sug66}. 
Therefore, pragmatic techniques are needed in an actual calculation~\cite{hee95, dug01}. 
We are going to focus on (iii).

In this study, we investigate the non-relativistic case and show that the Bogoliubov transformation, the particle-number parity, and 
the time-reversal symmetry in HFB are closely related with one another. 
From these findings, we can give a rigorous foundation for a method initiated by Bertsch et al.~\cite{ber09b}
to describe the 
ground state of odd-particle system 
as the lowest-energy state under an appropriate time-odd constraint in HFB theory. 
This method has a high affinity with DFT in the sense that
either an odd-particle system or an even-particle system 
is described as the ground state uniformly. 
We apply this method to the neutron-rich Mg isotopes near the drip line and 
demonstrate that it produces 
the exotic behavior in radii observed experimentally.

The article is organized as follows: 
In Sec.~\ref{theory}, after recapitulating the basics of HFB theory, the relationships among 
the Bogoliubov transformation, the particle-number parity, and 
the time-reversal symmetry are presented. 
In Sec.~\ref{method}, based on the relationships found in Sec.~\ref{theory}, 
we give foundation for a method describing the ground-state of odd-nuclei 
under a time-odd constraint. 
In Sec.~\ref{calculation}, we give a numerical procedure for describing the weakly-bound neutron-rich nuclei 
by employing the non-relativistic Skyrme-type EDF with the inclusion of 
the time-odd fields.
Then, results of the numerical analysis for Mg isotopes are presented. 
Finally, summary is given in Sec.~\ref{summary}. 

Part of the preliminary results of this work are reported in Ref.~\cite{kas20}. 

\section{Hartree--Fock--Bogoliubov theory for even and odd particle systems}\label{theory}

\subsection{Basics of HFB theory}
We begin with recalling the basics of Hartree--Fock--Bogoliubov (HFB) theory. 
The notation used here follows Ref.~\cite{rin80}. 
The Bogoliubov quasiparticle (qp) creation and annihilation operators $\hat{\beta}_k^\dagger, \hat{\beta}_k$ 
are defined as linear combinations of single-particle (sp) creation and annihilation operators $\hat{c}_k^\dagger, \hat{c}_k$:
\begin{equation}
\hat{\beta}_k^\dagger=\sum_{l} (U_{lk} \hat{c}_l^\dagger +V_{lk} \hat{c}_l),
\end{equation}
where indices $k$ and $l$ run over the whole configuration space ($k=1, \ldots, M$). 
Since we consider spin-$1/2$ particles, $M$ is an even number. 
The Bogoliubov transformation between the qp and sp bases is represented by the $2M\times 2M$ matrix 
\begin{equation}
	\label{eq:defW}
	\mathcal{W}\coloneqq\begin{pmatrix} U & V^* \\ V & U^* \end{pmatrix}
\end{equation}
as
\begin{equation}
	\begin{pmatrix} \hat{\beta} \\ \hat{\beta}^\dagger \end{pmatrix} 
	=\mathcal{W}^\dagger \begin{pmatrix} \hat{c} \\ \hat{c}^\dagger \end{pmatrix}.
\end{equation}
In order to satisfy the fermion anticommutation relations for quasiparticles, 
$\mathcal{W}$ must be unitary: $\mathcal{W}^\dagger \mathcal{W}=\mathcal{W} \mathcal{W}^\dagger=I_{2M}$
with $I_n$ representing the $n\times n$ identity matrix. 
The ground state wave function of the many-body system in HFB theory, or the HFB vacuum, 
$\ket{\Phi}$ is defined as the vacuum of the quasiparticles:
\begin{equation}
	\hat{\beta}_k \ket{\Phi} =0 \qquad \forall k. 
\end{equation}
The complete information about $\ket{\Phi}$ is contained by the density matrix 
$\rho_{kl}\coloneqq\braket{\Phi|\hat{c}_l^\dagger \hat{c}_k|\Phi} =(V^* V^\mathsf{T})_{kl}$ and pairing tensor $\kappa_{kl}\coloneqq\braket{\Phi|\hat{c}_l \hat{c}_k|\Phi} =(V^* U^\mathsf{T})_{kl}$, 
or by the generalized density matrix 
\begin{equation}
	\label{eq:defR}
	\mathcal{R}\coloneqq\begin{pmatrix} \rho & \kappa \\ -\kappa^*& 1-\rho^* \end{pmatrix}
	=\mathcal{W} \begin{pmatrix}O_M & O_M \\ O_M & I_M\end{pmatrix} \mathcal{W}^\dagger,
\end{equation}
where $O_n$ represents the $n\times n$ zero matrix. 
The unitarity of $\mathcal{W}$ guarantees that $\mathcal{R}$ is idempotent: $\mathcal{R}^2=\mathcal{R}$.
Under the idempotency of $\mathcal{R}$, 
the variational principle with a constraint on the expectation value of the particle number: $\delta \braket{\Phi |\hat{H}-\lambda \hat{N} |\Phi}=0$, 
where $\hat{H}$, $\hat{N}$, and $\lambda$ are 
the Hamiltonian of the system, the particle-number operator, and the chemical potential, respectively, 
leads $\mathcal{R}$ to commute with the HFB Hamiltonian
\begin{equation}
	\label{eq:HFBHamiltonian}
	\mathcal{H}
	\coloneqq
	\begin{pmatrix} h-\lambda I_M &\Delta \\ -\Delta^* & -h^*+\lambda I_M \end{pmatrix},
\end{equation}
where 
$h_{kl} \coloneqq \delta \braket{\Phi|\hat{H}|\Phi}/\delta \rho_{lk}$ and $\Delta_{kl} \coloneqq \delta \braket{\Phi|\hat{H}|\Phi}/\delta \kappa_{kl}^*$ are the sp and pair Hamiltonian, respectively. 
It follows that $\mathcal{R}$ and $\mathcal{H}$ are simultaneously diagonalized, 
and thus the HFB equations are represented in a matrix form as
\begin{equation}
	\begin{pmatrix} h-\lambda I_M &\Delta \\ -\Delta^* & -h^*+\lambda I_M \end{pmatrix} 
	\begin{pmatrix} U & V^* \\ V & U^* \end{pmatrix} 
	=
	\begin{pmatrix} U & V^* \\ V & U^* \end{pmatrix} 
	\begin{pmatrix} E & 0 \\ 0 & -E \end{pmatrix},
\end{equation}
or
\begin{equation}
	\mathcal{W}^\dagger \mathcal{H} \mathcal{W}= \mathcal{E}, \label{HFB}
\end{equation}
where $\mathcal{E}\coloneqq\mathrm{diag} (E, -E)$ and $E$ is a diagonal matrix of qp energies $E_k$. 
Note that the HFB Hamiltonian inherently has the particle-hole symmetry:
\begin{equation}
	\Sigma_x \mathcal{H}^* \Sigma_x=-\mathcal{H}, \label{eq_p-hsym}
\end{equation}
where
\[
	\Sigma_x 
	\coloneqq
	\begin{pmatrix} 0 & 1\\ 1& 0 \end{pmatrix} \otimes I_M = 
	\begin{pmatrix} O_M & I_M\\ I_M& O_M \end{pmatrix}. 
\]
It follows that when $\varphi_k$ is an eigenvector of $\mathcal{H}$ 
with eigenvalue $E_k$, $\Sigma_x \varphi_k^*$ is also an eigenvector with eigenvalue $-E_k$. 
Thus, the eigenvalues of $\mathcal{H}$ always come in pairs of opposite signs, 
but the theory says nothing about the individual signs of $E_k$. 
Therefore, we have to choose for each $k$ ($k=1, \cdots, M$) whether to take $E_k$ positive or negative. 
This choice for the solution of the HFB equations in superfluid systems corresponds to the choice of `occupied' or `unoccupied' orbits for the solution of the Hartree-Fock (HF) equations in normal systems.
A na\"ive choice to describe the ground state of the system is to take all $E_k$ positive 
as in the HF case the ground state is obtained by filling sp levels from the below.
Indeed, the state obtained by this choice has the lowest energy in the sense that 
all the qp excitations cost positive energies. 
As will be seen below, however, the state is not always the ground state of the system with desired particle-number parity~\cite{ban74}. 
We shall therefore call the state obtained by taking all $E_k$ positive the lowest-energy state, in distinction from other choices and the ground state with the proper particle-number parity. 


\subsection{Particle-number parity and the Bogoliubov transformation}
We first show a 
relationship between the particle-number parity $\pi_N$ and the Bogoliubov transformation matrix $\mathcal{W}$.
The famous theorem of Bloch and Messiah says that a unitary matrix $\mathcal{W}$ of the form (\ref{eq:defW})
can always be decomposed into three matrices of very special form~\cite{blo62}:
\begin{equation}
	\label{BM}
	\mathcal{W}
	=
	\begin{pmatrix} 
	D&0 \\ 0&D^* 
	\end{pmatrix} 
	\begin{pmatrix} 
	\bar{U}&\bar{V} \\ \bar{V}&\bar{U} 
	\end{pmatrix} 
	\begin{pmatrix} 
	C&0 \\ 0&C^* 
	\end{pmatrix}.
\end{equation}
Here $C$ and $D$ are unitary matrices, and $\bar{U}, \bar{V}$ are real matrices of the general form
\begin{align}
	\bar{U}=
	\begin{pmatrix}
	O_{N_1}&&&&&\\
	&\bar{U}^{(1)}&&&0&\\
	&&\bar{U}^{(2)}&&&\\
	&&&\ddots&&\\
	&0&&&\bar{U}^{(N_2)}&\\
	&&&&&I_{N_3}
	\end{pmatrix}, 
	&&\bar{V}=
	\begin{pmatrix}
	I_{N_1}&&&&&\\
	&\bar{V}^{(1)}&&&0&\\
	&&\bar{V}^{(1)}&&&\\
	&&&\ddots&&\\
	&0&&&\bar{V}^{(N_2)}&\\
	&&&&&O_{N_3}
	\end{pmatrix},
\end{align}
where $N_3=M-N_1-2N_2$, and $\bar{U}^{(p)}, \bar{V}^{(p)}$ are $2\times 2$ matrices of the form
\begin{equation}
	\bar{U}^{(p)}=\begin{pmatrix} u_p & 0 \\ 0 & u_p \end{pmatrix}, \quad \bar{V}^{(p)}=\begin{pmatrix} 0 & v_p \\ -v_p & 0 \end{pmatrix} \quad (p=1, 2, \dots, N_2),
\end{equation}
where $u_p$ and $v_p$ satisfy the conditions: $u_p > 0$, $v_p > 0$, $u_p^2+v_p^2=1$ ($p=1, 2, \dots, N_2$). 
One can explicitly construct the HFB vacuum $\ket{\Phi}$ in terms of the so called canonical basis 
defined as $\hat{a}_k^\dagger = \sum_l D_{lk} \hat{c}_l^\dagger$~\cite{rin80}:
\begin{equation}
	\ket{\Phi}=\prod_{i=1}^{N_1} \hat{a}_i^\dagger\prod_{p=1}^{N_2}(u_{p}+v_{p} \hat{a}_{p}^\dagger \hat{a}_{\bar{p}}^\dagger) \ket{0}. 
\end{equation}
Here $\ket{0}$ is the empty state defined as $\hat{c}_k \ket{0}= 0$ for all $k$. 
The index $\bar{p}$ represents an orbital paired with the orbital $p$, 
and $N_2$ indicates the maximum number of pairs in $\ket{\Phi}$. 
$N_1$ represents the number of unpaired particles, and corresponds to the seniority number in the quasi-spin theory \cite{ker61}. 
Depending on whether $N_1$ is even or odd, $\ket{\Phi}$ is a superposition of states with either even or odd particles, but never both. 
This means that the HFB vacuum $\ket{\Phi}$ is an eigenstate of the operator $\hat{P}_N=\e^{\ii\pi \hat{N}}$, 
where $\hat{N}$ is the particle-number operator. 
The eigenvalue $\pi_N=(-1)^{N_1}$ is a good quantum number, called the particle-number parity, or the number parity for short~\cite{ban73}. 
Note that the number parity has nothing to do with the average particle number $\braket{\Phi|\hat{N}|\Phi}$, which can be even, odd, or fractional, depending on the value of the chemical potential $\lambda$ in Eq.~(\ref{eq:HFBHamiltonian}).

Taking determinant of both sides of Eq.~(\ref{BM}), we obtain
\begin{align}
	\label{eq:determinant}
	\det \mathcal{W}
	&=
	\det 
	\begin{pmatrix} D&0 \\ 0&D^* \end{pmatrix} 
	\det \begin{pmatrix} \bar{U}&\bar{V} \\ \bar{V}&\bar{U} \end{pmatrix} 
	\det \begin{pmatrix} C&0 \\ 0&C^* \end{pmatrix} \notag \\
	&=
	\left|
	\det D\right|^2 
	\det (\bar{U}+\bar{V}) 
	\det (\bar{U}-\bar{V})  
	\left|\det C\right|^2 \notag \\
	&=
	\det 
	\left\{ 
	\diag 
	\left[ 
	I_{N_1}, 
	\begin{pmatrix} u_1 & v_1 \\ -v_1 & u_1 \end{pmatrix}, 
	\begin{pmatrix} u_2 & v_2 \\ -v_2 & u_2 \end{pmatrix}, 
	\cdots, 
	\begin{pmatrix} u_{N_2} & v_{N_2} \\ -v_{N_2} & u_{N_2} \end{pmatrix}, 
	I_{N_3}
	\right]
	\right. \notag \\
	&
	\hspace{1cm}
	\times
	\diag
	\left.
	\left[ 
	-I_{N_1}, 
	\begin{pmatrix} u_1 & -v_1 \\ v_1 & u_1 \end{pmatrix}, 
	\begin{pmatrix} u_2 & -v_2 \\ v_2 & u_2 \end{pmatrix}, 
	\cdots, 
	\begin{pmatrix} u_{N_2} & -v_{N_2} \\ v_{N_2} & u_{N_2} \end{pmatrix}, 
	I_{N_3}
	\right]
	\right\} \notag \\
	&=
	\det 
	\begin{pmatrix} -I_{N_1}& 0 \\ 0 & I_{M-N_1} \end{pmatrix}
	=(-1)^{N_1},
\end{align}
where we used the fact that $C$ and $D$ are unitary matrices, 
and 
\[
	\begin{pmatrix} u_p & v_p \\ -v_p & u_p \end{pmatrix} 
	\begin{pmatrix} u_p & -v_p \\ v_p & u_p \end{pmatrix} 
	=
	\begin{pmatrix} u_p^2+v_p^2 & 0 \\ 0 & u_p^2+v_p^2 \end{pmatrix}
	=
	\begin{pmatrix} 1 & 0 \\ 0 & 1 \end{pmatrix}
	\qquad 
	(p=1, 2, \dots, N_2) .
\]
Eq.~(\ref{eq:determinant}) shows that $\det \mathcal{W}$ is nothing less than the number parity:
\begin{equation}
	\label{eq:numberparity}
	\pi_N=\det \mathcal{W}. 
\end{equation}

Before going ahead, we give a useful formula that relates the number parity $\pi_N$ to the HFB Hamiltonian $\mathcal{H}$ and the qp energies $E_k$. 
The fact that the determinant of $\mathcal{W}$ is either $+1$ or $-1$ 
implies that $\mathcal{W}$ can be unitarily transformed into an orthogonal matrix. 
In fact, by use of a unitary matrix
\begin{equation}
	\label{eq:X}
	\mathcal{X}=\frac{1}{\sqrt{2}}\begin{pmatrix}1&1\\\ii&-\ii\end{pmatrix}\otimes I_M,
\end{equation}
$\mathcal{W}$ is transformed into a real orthogonal matrix:
\begin{equation}
	\mathcal{W}_\mathcal{X}
	\coloneqq 
	\mathcal{X}\mathcal{W}\mathcal{X}^\dagger
	=
	\begin{pmatrix}
	\Re (U+V) & \Im (U+V) \\
	-\Im (U-V) & \Re (U-V)
	\end{pmatrix}.
\end{equation}
In the same way, the HFB Hamiltonian $\mathcal{H}$ is transformed into a pure-imaginary skew-symmetric matrix:
\begin{equation}
	\mathcal{H}_\mathcal{X}
	\coloneqq 
	\mathcal{X}\mathcal{H}\mathcal{X}^\dagger
	=
	\ii
	\begin{pmatrix}
	\Im (h'+\Delta) & - \Re (h'-\Delta) \\
	\Re (h'+\Delta) & \Im (h'-\Delta)
	\end{pmatrix},
\end{equation}
where $h' \coloneqq h-\lambda I_M$.
Then the HFB equations (\ref{HFB}) are rewritten as follows:
\begin{equation}
\mathcal{W}_\mathcal{X}^\mathsf{T}\mathcal{H}_\mathcal{X}\mathcal{W}_\mathcal{X}=\mathcal{E}_\mathcal{X},\label{HFBX}
\end{equation}
where 
\begin{equation}
\mathcal{E}_\mathcal{X}\coloneqq \mathcal{X}\mathcal{E}\mathcal{X}^\dagger=-\ii \begin{pmatrix}0&E\\-E&0\end{pmatrix}.
\end{equation}
Since the both sides of Eq.~(\ref{HFBX}) are $2M\times 2M$ skew-symmetric matrices, we can take Pfaffian\footnote{
	For a $2n \times 2n$ skew-symmetric matrix $A$ with matrix elements $a_{ij}$, the Pfaffian of $A$ is defined as
	\[
		\pf (A) 
		= 
		\frac{1}{2^n n!} 
		\sum_{\sigma \in S_{2n}} 
		\sgn (\sigma) 
		a_{\sigma (1) \sigma(2)} 
		a_{\sigma (3) \sigma(4)} 
		\dots 
		a_{\sigma (2n-1) \sigma (2n)},
	\]
	where $S_{2n}$ is the set of permutations on $2n$ elements.
	Pfaffians have the following properties: 
	For a $2n \times 2n$ skew-symmetric matrix $A$,
	\[
		(\pf A)^2=\det A.
	\]
	For a $2n \times 2n$ skew-symmetric matrix $A$ and an arbitrary $2n\times 2n$ matrix $B$,
	\[
		\pf (B^\mathsf{T} A B)=\det B \pf A.
	\]
	For an arbitrary $n \times n$ matrix $C$,
	\[
		\pf \begin{pmatrix}0&C\\-C^\mathsf{T}&0\end{pmatrix}=(-1)^{n(n-1)/2} \det C.
	\]
	For the proof of these properties and more details on the Pfaffian, see, e.g., Ref.~\cite{hab15} and the appendix of Ref.~\cite{rob09}.
} 
of the both sides. 
By use of the properties of the Pfaffian: $\pf (\mathcal{W}_\mathcal{X}^\mathsf{T}\mathcal{H}_\mathcal{X}\mathcal{W}_\mathcal{X}) = \det \mathcal{W}_\mathcal{X} \pf \mathcal{H}_\mathcal{X}$ and $\pf \mathcal{E}_\mathcal{X} = (-1)^{M(M-1)/2} \det (-\ii E)$, we obtain
\[
	\det \mathcal{W}_\mathcal{X} 
	\pf \mathcal{H}_\mathcal{X}
	=
	(-1)^{M(M-1)/2} 
	\det (-\ii E).
\]
Since $M$ is an even number, $E = \diag (E_1, E_2, \dots, E_M)$, and $\det \mathcal{W}_\mathcal{X} = \det \mathcal{W}$, it follows
\begin{equation}
	\label{eq:formula}
	\det \mathcal{W}
	\pf \mathcal{H}_\mathcal{X}
	=
	\prod_{k=1}^M E_k.
\end{equation}
Thus we arrive at a useful relation between the number parity $\pi_N$, the HFB Hamiltonian $\mathcal{H}$, and the qp energies $E_k$: 
\begin{equation}
	\label{eq:formula2}
	\pi_N
	\pf \mathcal{H}_\mathcal{X}
	=
	\prod_{k=1}^M E_k.
\end{equation}
We emphasize that Eq.~(\ref{eq:formula2}) is a general relation that holds regardless of the signs of $E_k$. 
For the lowest-energy state, the right-hand side of Eq.~(\ref{eq:formula2}) is positive by definition. 
Therefore, the number parity of the lowest-energy state, $\pi_N^{\rm L.E.}$, depends only on the sign of $\pf \mathcal{H}_{\mathcal{X}}$:
\begin{equation}
	\label{eq:formulaLE}
	\pi_N^{\rm L.E.} = \sgn \pf \mathcal{H}_{\mathcal{X}}.
\end{equation}
Note that an equivalent formula to Eq.~(\ref{eq:formulaLE}) is obtained in Ref.~\cite{kit01} to investigate topological properties of a one-dimensional superfluid system.

\subsection{Symmetries in HFB theory}\label{subsec:symmetry}
Before investigating the time-reversal symmetry of the HFB Hamiltonian, we are going to give 
a general property of the symmetry in HFB theory. 
Since the HFB equations are nonlinear, the HFB Hamiltonian $\mathcal{H}$ does not necessarily hold the same symmetries 
as the Hamiltonian of the system $\hat{H}$. 
Nevertheless, certain symmetries are still conserved in HFB theory. 
Such symmetries are called self-consistent symmetries, 
and they often significantly reduce the dimension of the eigenvalue problem~\cite{goo74}.

Consider a symmetry transformation realized by 
a unitary or anti-unitary operator $\hat{U}_s$ which maps the sp space into itself by a $M\times M$ unitary matrix $U_s$:
\begin{equation}
	\hat{U}_s^\dagger \hat{c}_k \hat{U}_s=\sum_l U_{s\, kl} \hat{c}_l, 
\end{equation}
or in the $2M$-dimensional space
\begin{equation}
	\hat{U}_s^\dagger \begin{pmatrix} \hat{c} \\ \hat{c}^\dagger \end{pmatrix} \hat{U}_s=\begin{pmatrix} U_s & 0 \\ 0 & U_s^* \end{pmatrix} \begin{pmatrix} \hat{c} \\ \hat{c}^\dagger \end{pmatrix} = \mathcal{U}_s \begin{pmatrix} \hat{c} \\ \hat{c}^\dagger \end{pmatrix},
\end{equation}
where $\mathcal{U}_s \coloneqq \mathrm{diag} (U_s, U_s^*)$ is a $2M\times 2M$ unitary matrix. 
Under the transformation $\ket{\Phi} \rightarrow \hat{U}_s \ket{\Phi}$, 
the generalized density matrix $\mathcal{R}$ transforms as
\footnote{Note the property of an anti-unitary operator $\hat{\Theta}$: $(\bra{\Phi} \hat{\Theta}^\dagger) \, \hat{O}\, (\hat{\Theta} \ket{\Phi})=(\braket{\Phi|(\hat{\Theta}^\dagger \hat{O} \hat{\Theta})|\Phi})^*$.}
\begin{equation}
	\label{eq:transR}
	\mathcal{R} \rightarrow (\mathcal{U}_s \mathcal{R} \mathcal{U}_s^\dagger)^{(*)}.
\end{equation}
Here $(\cdots)^{(*)}$ denotes that the complex conjugate is taken if $\hat{U}_s$ is an anti-unitary operator. 
Assuming that $\hat{U}_s$ is a symmetry operator of the system, that is $[\hat{H}, \hat{U}_s]=0$, 
we find the HFB Hamiltonian $\mathcal{H}$ transforms in the same way as the generalized density matrix $\mathcal{R}$:
\begin{equation}
	\label{eq:transH}
	\mathcal{H} \rightarrow (\mathcal{U}_s \mathcal{H} \mathcal{U}_s^\dagger)^{(*)}.
\end{equation}
Now suppose that the HFB vacuum $\ket{\Phi}$ is invariant up to a phase under the operation $\hat{U}_s$, 
i.e. $\hat{U}_s$ is a symmetry operator of the intrinsic system, 
it follows that
\begin{equation}
	\mathcal{H} = (\mathcal{U}_s \mathcal{H} \mathcal{U}_s^\dagger)^{(*)}. 
	\label{eq:SCS}
\end{equation}
When $\hat{U}_s$ is a unitary operator, it leads to $[\mathcal{H}, \mathcal{U}_s]=0$. 
This indicates that the HFB Hamiltonian $\mathcal{H}$ is block diagonalized with respect to the conserved quantum numbers 
associated with the transformation $\hat{U}_s$.

In particular, we are interested in such $\hat{U}_s$ that is generated by a Hermitian 
particle-hole one-particle operator $\hat{S}=\sum_{kl} S_{kl} \hat{c}^\dagger_k \hat{c}_l$: $\hat{U}_s=\e^{\ii\theta \hat{S}}$ 
or $\e^{\ii\theta \hat{S}}\hat{K}$, 
where $\theta$ is a real parameter  and $\hat{K}$ is the complex conjugation operator 
which leaves the sp basis $\ket{k}=c^\dagger_k\ket{0}$ invariant: $\hat{K} \ket{k} = \ket{k}$. 
In this case, $U_s=\e^{\ii\theta S}$. 
Especially when $\hat{U}_s=\e^{\ii\theta \hat{S}}$,  Eq.~(\ref{eq:SCS}) leads to 
\begin{equation}
\left[ \mathcal{H}, \begin{pmatrix} \e^{\ii\theta S}&0 \\ 0& \e^{-\ii\theta S^*} \end{pmatrix} \right]=0.
\label{eq:SCS2}
\end{equation}
If this holds for an arbitrary $\theta$, one sees 
\begin{equation}
\left[ \mathcal{H}, \begin{pmatrix} S&0 \\ 0& -S^* \end{pmatrix} \right]=0.
\label{eq:SCS3}
\end{equation}
Note that since the HFB vacuum is always an eigenstate of the number parity operator $\hat{P}_N=\e^{\ii\pi\hat{N}}$, Eq.~(\ref{eq:SCS2})  reduces to a trivial commutation relation $[\mathcal{H}, -I_{2M}]=0$ when $\hat{S}=\hat{N}$ and $\theta=\pi$.

The HFB equations achieve self-consistency between the densities and the potentials by an iterative process. 
We are going to discuss here whether the intrinsic symmetry defined in Eq.~(\ref{eq:SCS}) is affected by the iterations. 
From Eqs.~(\ref{eq:transR}) and (\ref{eq:transH}) the HFB Hamiltonian has the following property as a functional of the generalized density matrix $\mathcal{R}$ for a symmetry $\hat{U}_s$ of the system as in the case of HF: 
\begin{equation}
	\label{eq:sympro}
	\left(
	\mathcal{U}_s 
	\mathcal{H}[\mathcal{R}]
	\mathcal{U}_s^\dagger 
	\right)^{(*)}
	=
	\mathcal{H}
	\left[(
	\mathcal{U}_s 
	\mathcal{R}
	\mathcal{U}_s^\dagger 
	)^{(*)}\right].
\end{equation}
This means that if the initial density $\mathcal{R}^{(0)}$ has a certain symmetry, then the mean-field Hamiltonian $\mathcal{H}[\mathcal{R}^{(0)}]$ for the first step of the iteration has it too. The density $\mathcal{R}^{(1)}$ for the next step is found by diagonalization of $\mathcal{H}[\mathcal{R}^{(0)}]$, hence the same symmetry holds. In each step of the iteration, the intrinsic symmetry is thus conserved.
Note that the average particle number is fixed at the desired value by adjusting the chemical potential $\lambda$ in $\mathcal{H}$ throughout the iterations.

\subsection{Time-reversal symmetry and number parity}
Using the results obtained above, 
we show that the time-reversal symmetry of the HFB Hamiltonian and the number parity of the HFB vacuum are directly related 
to each other.

Let us consider the case when the HFB Hamiltonian has time-reversal symmetry. 
Paying attention to the anti-unitarity of the time-reversal operator 
$\hat{T}=\mathrm{exp}(-\ii\pi\hat{S}_y) \hat{K}$, 
where $\hat{S}_y$ is the $y$-component of the total spin operator, we have the following equality relation from Eq.~(\ref{eq:SCS})
\begin{equation}
	\label{eq_trsym}
	\mathcal{H}=\mathcal{T}\mathcal{H}^*\mathcal{T}^\mathsf{T},
\end{equation}
where
\begin{equation}
	\mathcal{T}
	\coloneqq 
	\begin{pmatrix}T & 0 \\ 0 & T \end{pmatrix}, 
	\qquad 
	T 
	\coloneqq 
	e^{-\ii\pi S_y} 
	= 
	I_{M/2}\otimes -\ii\sigma_y 
	=
	\begin{pmatrix}
	0&-1&&&&&\\ 
	1&0&&&&0&\\ 
	&&0&-1&&&\\ 
	&&1&0&&&\\ 
	&&&&\ddots&& \\ 
	&0&&&&0&-1 \\ 
	&&&&&1&0 
	\end{pmatrix}.
\end{equation}
It follows that when $\varphi_k$ is an eigenvector of $\mathcal{H}$ with eigenvalue $E_k$, 
the time-reversed state $\mathcal{T}\varphi_k^*$ is an independent eigenvector of $\mathcal{H}$ 
with the same eigenvalue $E_k$. In other words, $\varphi_k$ and $\mathcal{T}\varphi_k^*$ are degenerated, 
i.e. the Kramers degeneracy shows up. 
Since the particle-hole symmetry (\ref{eq_p-hsym}) is always kept in HFB theory, 
$\Sigma_x \varphi_k^*$ and $\Sigma_x \mathcal{T} \varphi_k$ are also Kramers-degenerated eigenvectors with eigenvalue $-E_k$. 
Therefore, 
a specific solution of the HFB equations is constructed as follows:
\begin{align}
	&\widetilde{\mathcal{W}}
	=
	\begin{pmatrix} \widetilde{U} & \widetilde{V}^* \\ \widetilde{V} & \widetilde{U}^* \end{pmatrix}, \\
	\label{eq:Utilde}
	&\widetilde{U}
	=
	\begin{pmatrix} u_1 & Tu_1^* & u_2 & Tu_2^* & \cdots & u_{M/2} & Tu_{M/2}^* \end{pmatrix}, \\
	&
	\widetilde{V}
	=
	\begin{pmatrix} v_1 & Tv_1^* & v_2 & Tv_2^* & \cdots & v_{M/2} & Tv_{M/2}^* \end{pmatrix},
\end{align}
where $\varphi_k = \bigl( \begin{smallmatrix} u_k \\ v_k \end{smallmatrix}\bigr)$, $\mathcal{T} \varphi_k^* = \bigl( \begin{smallmatrix} T u_k^* \\ T v_k^* \end{smallmatrix}\bigr)$, $\Sigma_x \varphi_k^* = \bigl( \begin{smallmatrix} v_k^* \\ u_k^* \end{smallmatrix}\bigr)$, and $\Sigma_x \mathcal{T} \varphi_k = \bigl( \begin{smallmatrix} T v_k \\ T u_k \end{smallmatrix}\bigr)$ are orthonormal eigenvectors of $\mathcal{H}$ with eigenvalues $E_k$, $E_k$, $-E_k$, and $-E_k$, respectively. 
Here, we take all $E_k$ non-negative without loss of generality. 
The orthogonality of $\varphi_k$ and $\mathcal{T} \varphi_k^*$ is ensured as $\varphi_k^\dagger \mathcal{T} \varphi_k^* = (\varphi_k^\dagger \mathcal{T} \varphi_k^*)^\mathsf{T} = \varphi_k^\dagger \mathcal{T}^\mathsf{T} \varphi_k^* = -\varphi_k^\dagger \mathcal{T} \varphi_k^* = 0$. If $E_k \neq 0$, $\varphi_k$ is orthogonal to $\Sigma_x \varphi_k^*$ because they are eigenvectors of the same Hermitian 
matrix with different eigenvalues, and even if $E_k = 0$, it is possible to redefine $\varphi_k$ to be orthogonal to $\Sigma_x \varphi_k^*$ by mixing $\varphi_k$ and $\mathcal{T} \varphi_k^*$. Thus, $\widetilde{\mathcal{W}}$ is a unitary matrix. 
In addition, we find $\widetilde{\mathcal{W}}$ is a symplectic matrix, that is, it satisfies
\begin{equation}
	\label{eq:symplectic}
	\widetilde{\mathcal{W}}^\mathsf{T} 
	\mathcal{T} 
	\widetilde{\mathcal{W}}
	=
	\mathcal{T}.
\end{equation}
This is shown as follows: 
Multiplying both sides of Eq.~(\ref{eq:Utilde}) by $T$ from the left and $T^\mathsf{T}$ from the right, we get
\begin{align*}
	T\widetilde{U}T^\mathsf{T}
	&=
	T \begin{pmatrix} u_1 & Tu_1^* & u_2 & Tu_2^* & \cdots & u_{M/2} & Tu_{M/2}^* \end{pmatrix} T^\mathsf{T} \notag \\
	&=
	\begin{pmatrix} Tu_1 & -u_1^* & Tu_2 & -u_2^* & \cdots & Tu_{M/2} & -u_{M/2}^* \end{pmatrix} T^\mathsf{T} \notag \\
	&=
	\begin{pmatrix} u_1^* & Tu_1 & u_2^* & Tu_2 & \cdots & u_{M/2}^* & Tu_{M/2} \end{pmatrix} =\widetilde{U}^*.
\end{align*}
In the same way, one obtains $T\widetilde{V}T^\mathsf{T}=\widetilde{V}^*$. Thus, it follows that
\begin{equation}
	\label{eq:Wtilde}
	\mathcal{T} \widetilde{\mathcal{W}}\mathcal{T}^\mathsf{T}=\widetilde{\mathcal{W}}^*.
\end{equation}
Then, multiplying the both sides of Eq.~(\ref{eq:Wtilde}) by $\widetilde{\mathcal{W}}^\mathsf{T}$ from the left and $\mathcal{T}$ from the right, and using the fact that $\widetilde{\mathcal{W}}$ is a unitary matrix and $\mathcal{T}$ is an orthogonal matrix, we obtain Eq.~(\ref{eq:symplectic}).
It is known that the determinant of any symplectic matrix is $+1$. 
This is easily shown through the use of Pfaffian: Taking Pfaffian of the both sides of Eq.~(\ref{eq:symplectic}), it follows that
$
\det \widetilde{\mathcal{W}} \, \mathrm{pf}\, \mathcal{T} = \mathrm{pf}\, \mathcal{T} 
$, 
and since $\mathrm{pf}\, \mathcal{T} =(-1)^M \neq 0$, one sees
\begin{equation}
	\label{eq:detWtilde}
	\det \widetilde{\mathcal{W}} =+1.
\end{equation}
Now the HFB equations are written as follows: 
\begin{equation}
	\label{eq:HFBWtilde}
	\widetilde{\mathcal{W}}^\dagger 
	\mathcal{H} 
	\widetilde{\mathcal{W}} 
	= 
	\widetilde{\mathcal{E}},
\end{equation}
where $\widetilde{\mathcal{E}}=\diag (\tilde{E}, -\tilde{E})$ and $\tilde{E}=\diag (E_1, E_1, E_2, E_2, \dots, E_{M/2}, E_{M/2})$. 
From Eqs.~(\ref{eq:formula}) and (\ref{eq:detWtilde}) we have
\begin{equation}
	\label{eq:pfH}
	\pf \mathcal{H}_{\mathcal{X}} = \prod_{k=1}^{M/2} E_k^2.
\end{equation}
It follows that $\pf \mathcal{H}_{\mathcal{X}} \geq 0$ with equality if $E_k=0$ for at least one $k$. 
This together with Eq.~(\ref{eq:formulaLE}) means that the lowest-energy solution of the HFB equations has the even number parity as long as the time-reversal symmetry is conserved in the HFB Hamiltonian with no null eigenvalues.

Next we shall show that the lowest-energy state is not well-defined when the HFB Hamiltonian has null eigenvalues. A general solution of the HFB equations under the time-reversal symmetry is obtained by
\begin{equation}
	\label{eq:generalW}
	\mathcal{\mathcal{W}}
	=
	\widetilde{\mathcal{W}}
	\mathcal{Z},
\end{equation}
where $\mathcal{Z}$ is a $2M \times 2M$ unitary matrix 
which consists of the unitary transformations containing the permutation of columns as well as the phase transformation and mixing of the degenerate qp states. 
From Eqs.~(\ref{eq:numberparity}), (\ref{eq:detWtilde}) and (\ref{eq:generalW}), 
the number parity of the state under consideration is now given by $\pi_N = \det \mathcal{Z}$. 
Only the permutation of the $k$th and $(M+k)$th columns ($k \leq M$), that is, a quasi-particle state with eigenvalue $E_k$ and the corresponding quasi-hole state with eigenvalue $-E_k$, changes the number parity.
This is because for $\mathcal{W}$ to be written in the form (\ref{eq:defW}) the permutation of the $k$th and $l$th columns ($k < l \leq M$) accompany the permutation of the $(M+k)$th and $(M+l)$th columns, 
and no other permutations except that of the $k$th and $(M+k)$th columns are allowed. 
For the lowest-energy state, however, the permutation of the quasi-particle and quasi-hole states is permitted only if $E_k=0$ because the permutation means swapping of $E_k$ and $-E_k$ and goes against the definition of the lowest-energy state unless $E_k = 0$. 
When $E_k=0$, any mixing of $\varphi_k$, $\mathcal{T}\varphi_k^*$, $\Sigma_x\varphi_k^*$, and $\Sigma_x\mathcal{T}\varphi_k$, or, the $k$th, $(k+1)$th, $(M+k)$th, and $(M+k+1)$th columns, is allowed for the lowest-energy state since they are all degenerated, and such a mixing can change the number parity of the lowest-energy state as discussed above. Therefore, we cannot uniquely determine the lowest-energy state with a specific number parity when the HFB Hamiltonian has null eigenvalues. 

Finally, we would like to make sure that the above conclusion is not broken via non-linear effects. 
If the HFB Hamiltonian does not have null eigenvalues, there is no mixing between quasi-particle and quasi-hole states for the lowest-energy state. Thus, the lowest-energy solution $\mathcal{W}_{\mathrm{L.E.}}$ is written as
\begin{equation}
	\label{eq:constructWLE}
	\mathcal{W}_{\mathrm{L.E.}} = \widetilde{\mathcal{W}} \begin{pmatrix} X & 0 \\ 0 & X^* \end{pmatrix},
\end{equation}
where $X$ is a $M \times M$ unitary matrix. 
Then $\det \mathcal{W}_{\mathrm{L.E.}} = \left| \det X \right|^2=+1$, and we get the above conclusion again: the lowest-energy solution of the HFB equation has the even number parity as long as the time-reversal symmetry is conserved in the HFB Hamiltonian without null eigenvalues. 
From Eq.~(\ref{eq:defR}), the generalized density matrix for the lowest-energy state is obtained by
\begin{equation}
	\mathcal{R}_{\mathrm{L.E.}} 
	= 
	\mathcal{W}_{\mathrm{L.E.}} 
	\begin{pmatrix} O_M & O_M \\ O_M & I_M \end{pmatrix}
	\mathcal{W}_{\mathrm{L.E.}}^\dagger
	=
	\widetilde{\mathcal{W}}
	\begin{pmatrix} O_M & O_M \\ O_M & I_M \end{pmatrix}
	\widetilde{\mathcal{W}}^\dagger.
\end{equation}
From Eq.~{(\ref{eq:Wtilde})}, this $\mathcal{R}_{\mathrm{L.E.}}$ has the time-reversal symmetry:
\begin{equation}
	\mathcal{T} \mathcal{R}_{\mathrm{L.E.}}^* \mathcal{T}^\mathsf{T} = \mathcal{R}_{\mathrm{L.E.}}.
\end{equation}
Therefore, 
from the discussion in Sec.~\ref{subsec:symmetry}, the lowest-energy state constructed as Eq.~{(\ref{eq:constructWLE})} has even-number parity at each step of the iteration.

\section{Methodology for describing odd-mass nuclei}\label{method}
We demonstrated above that 
the lowest-energy state is always an even number-parity state, that is, an even particle system
in HFB theory as long as the time-reversal symmetry is conserved for the intrinsic Hamiltonian. 
The procedure called the blocking method, which has been conventionally used to describe an odd number-parity state~\cite{sug66, rin74, sch10}, 
can be viewed as follows: 
One solves the time-reversal symmetric HFB equations to generate a reference state with $\det \mathcal{W} =+1$, 
and then swaps one set of columns of $\mathcal{W}$ to obtain a state whose sign of determinant is reversed.
Alternatively, one can take a strategy to describe an odd number-parity state 
as the lowest-energy state under the constraint which breaks time-reversal symmetry. 
The difference between our method and the blocking method is explained in Eq.~(\ref{eq:formula2}). 
In the blocking method, to change the number parity $\pi_N$ one replaces $E_k$ by $-E_k$ for a certain $k$ in the right-hand side of Eq.~(\ref{eq:formula2}) leaving $\pf \mathcal{H}_{\mathcal{X}}$ as it is.
In our method, on the other hand, we change the sign of $\pf \mathcal{H}_{\mathcal{X}}$ by a time-odd constraint while keeping the sign of the right-hand side to change $\pi_N$. 
The idea of obtaining odd-mass nuclei by such a constraint was proposed by Bertsch et al., 
stimulated by the non-collective cranking method~\cite{ber09b}. 
In what follows, 
we generalize this approach and encapsulate the essential point of this method. 

Assuming that the intrinsic system is invariant under any unitary transformation 
generated by a Hermitian time-odd particle-hole type one-body operator 
$\hat{S}=\sum_{ij}S_{ij}c_i^\dagger c_j$ with $S^\dagger =S$, and $T S^* T^\mathsf{T}=-S$. 
For example, the $z$-component of the total angular momentum $\hat{J}_z$ can be employed as $\hat{S}$ for an axially-symmetric system around the $z$-axis. 
Since $\hat{S}$ is a symmetry of the intrinsic system, the mean-field representation of $\hat{S}$ commutes with the HFB Hamiltonian $\mathcal{H}$ [see Eq.~(\ref{eq:SCS3})]:
\begin{equation}
[\mathcal{H}, \mathcal{S}]=0, \label{eq_commutation}
\end{equation}
where
\begin{equation}
\mathcal{S}=\begin{pmatrix} S&0 \\ 0&-S^* \end{pmatrix}.
\end{equation}
Introducing a Lagrange multiplier $\lambda_s$ to fix the expectation value of $\hat{S}$ 
besides the chemical potential $\lambda$, we consider the variational principle
\begin{equation}
\delta\braket{\hat{H}-\lambda\hat{N}-\lambda_s\hat{S}}=0.
\end{equation}
It then gives the HFB Routhian 
\begin{equation}
\mathcal{H}'=\mathcal{H}-\lambda_s \mathcal{S}.
\end{equation}
From the commutation relation (\ref{eq_commutation}), 
$\mathcal{H}$ can be block diagonalized for each eigenvalue of $\mathcal{S}$. 
Since $\mathcal{S}$ is proportional to the identity matrix in each block, 
the eigenvalues of $\mathcal{H}'$ are linearly shifted from those of $\mathcal{H}$. 
Then, as shown below, 
the sign of $\mathrm{pf}\, \mathcal{H}'_\mathcal{X}$ can change according to $\lambda_s$
and thus the number parity of the lowest-energy state can vary from positive to negative. 

Because of the time-odd character of $\hat{S}$, the eigenvalues of $S$ always come in pairs of opposite signs. 
Let $\{x^{\pm\alpha}_n\}_{n=1,2,\cdots}$ be sets of orthonormalized eigenvectors of $S$ 
with eigenvalue $\pm \omega_\alpha$ ($\alpha>0$, $\omega_\alpha>0$), 
where $n$ is a label that distinguishes states other than $\alpha$. Then
\begin{equation}
\left\{\begin{pmatrix}x^\alpha_n \\ 0\end{pmatrix}, \ \begin{pmatrix}0 \\ x^{-\alpha*}_n\end{pmatrix}\right\}_{n=1,2,\cdots}
\end{equation}
is a set of orthonormalized eigenvectors of $\mathcal{S}$ with eigenvalue $\omega_\alpha$. An appropriate linear combination
\begin{equation}
\chi^\alpha_n =\sum_{m} \left[U^\alpha_{mn} \begin{pmatrix}x^\alpha_m \\ 0\end{pmatrix} + V^\alpha_{mn} \begin{pmatrix}0 \\ x^{-\alpha*}_m\end{pmatrix}\right]
\end{equation}
is a simultaneous eigenstate of $\mathcal{S}$ and $\mathcal{H}$ 
with an eigenvalue $\omega_\alpha$ and $E^\alpha_n$, respectively. 
Thanks to the particle-hole symmetry of $\mathcal{H}$,
\begin{equation}
\varphi^\alpha_n \coloneqq \Sigma_x \left(\chi^{-\alpha}_n\right)^* =\sum_{m} \left[V^{-\alpha*}_{mn} \begin{pmatrix}x^\alpha_m \\ 0\end{pmatrix} + U^{-\alpha*}_{mn} \begin{pmatrix}0 \\ x^{-\alpha*}_m\end{pmatrix}\right]
\end{equation}
is also a simultaneous eigenstate of $\mathcal{S}$ and $\mathcal{H}$ 
with an eigenvalue $\omega_\alpha$ and $-E^{-\alpha}_n$, respectively. 
Therefore, the HFB equations are block diagonalized for each eigenvalue of $\mathcal{S}$ as follows:
\begin{equation}
\mathcal{W}^{\alpha \dagger} \mathcal{H}^\alpha \mathcal{W}^\alpha = \mathcal{E}^\alpha,
\end{equation}
where
\begin{align}
\mathcal{H}^\alpha=\begin{pmatrix} (h-\lambda I)^\alpha &\Delta^\alpha \\ -\Delta^{-\alpha*} & -(h-\lambda I)^{-\alpha*} \end{pmatrix}, \qquad \mathcal{W}^\alpha=\begin{pmatrix} U^\alpha & V^{-\alpha*} \\ V^\alpha & U^{-\alpha*} \end{pmatrix}, \qquad \mathcal{E}^\alpha=\begin{pmatrix} E^\alpha & 0 \\ 0 & -E^{-\alpha} \end{pmatrix},
\end{align}
and
\begin{equation}
(h -\lambda I)^\alpha_{mn}= x^{\alpha\dagger}_m (h-\lambda I_{M}) x^\alpha_n, \qquad \Delta^\alpha_{mn}= x^{\alpha\dagger}_m \Delta x^{-\alpha*}_n, \qquad E^\alpha=\mathrm{diag} (E^\alpha_1, E^\alpha_2, \cdots ).
\end{equation}
Since $\chi^\alpha_n$ and $\varphi^\alpha_n$ are simultaneous eigenstates of $\mathcal{S}$ and $\mathcal{H}$, 
they are also eigenstates of $\mathcal{H}'$ 
with an eigenvalue $E^\alpha_n -\lambda_s \omega_\alpha$ and $-E^{-\alpha}_n -\lambda_s \omega_\alpha$, 
respectively. In other words,
\begin{equation}
\mathcal{W}^{\alpha \dagger} \mathcal{H}'^\alpha  \mathcal{W}^\alpha = \mathcal{E}^\alpha - \lambda_s \omega_\alpha I^\alpha,
\end{equation}
where $\mathcal{H}'^\alpha=\mathcal{H}^\alpha -\lambda_s \omega_\alpha I^\alpha$,
and $I^\alpha$ is the identity matrix for the block with eigenvalue $\omega_\alpha$. 
In this way, one sees that 
the constraint on the intrinsic symmetry $\hat{S}$ does not change individual single qp states, 
but shifts only the qp energies according to the eigenvalues of $\mathcal{S}$. 
Reflecting the time-odd character of $\hat{S}$, the qp energies of time-reversal pairs split in the opposite direction. 
Therefore, even if the original HFB Hamiltonian $\mathcal{H}$ has time-reversal symmetry, 
the Kramers degeneracy is resolved at $\lambda_s \neq 0$.

Now, we are going to show that the number parity of the lowest-energy state 
can change according to $\lambda_s$.
The following unitary matrix diagonalizes $\mathcal{H}$ and $\mathcal{H}'$ simultaneously:
\begin{align}
&\underline{\mathcal{W}}=\begin{pmatrix} \underline{U} &\underline{V}^* \\ \underline{V} &\underline{U}^* \end{pmatrix},\\
&\underline{U}=\begin{pmatrix} \underline{U}^1& \underline{U}^{-1} & \underline{U}^2 & \underline{U}^{-2} &\cdots \end{pmatrix}, \qquad \underline{U}^\alpha_n=\sum_{m} x^\alpha_m U^\alpha_{mn}, \\
&\underline{V}=\begin{pmatrix} \underline{V}^1& \underline{V}^{-1} & \underline{V}^2 & \underline{V}^{-2} &\cdots \end{pmatrix}, \qquad \underline{V}^\alpha_n=\sum_{m} x^{-\alpha*}_m V^\alpha_{mn}.
\end{align}
Then the HFB equations for $\mathcal{H}$ and $\mathcal{H}'$ read
\begin{align}
	\underline{\mathcal{W}}^\dagger 
	\mathcal{H} 
	\underline{\mathcal{W}}
	=
	\underline{\mathcal{E}}, 
	\qquad
	\underline{\mathcal{W}}^\dagger 
	\mathcal{H}' 
	\underline{\mathcal{W}}
	=
	\underline{\mathcal{E}}',
\end{align}
where $\underline{\mathcal{E}}=\mathrm{diag}\, (\underline{E}, -\underline{E})$, $\underline{E}=\mathrm{diag}\, (E^1, E^{-1}, \cdots)$, and $\underline{\mathcal{E}}'=\mathrm{diag}\, (\underline{E}', -\underline{E}')$, $\underline{E}'=\mathrm{diag}\, (E^1-\lambda_s\omega_1I^1, E^{-1} +\lambda_s\omega_1I^1, \cdots)$. 
From Eq.~(\ref{eq:formula}) one obtains 
\begin{align}
	\label{eq_m1}
	&\det \underline{\mathcal{W}} 
	\pf \mathcal{H}_\mathcal{X}
	=
	\prod_{n, \alpha>0} 
	E^\alpha_n E^{-\alpha}_n, \\
	\label{eq_m2}
	&\det \underline{\mathcal{W}} 
	\pf \mathcal{H}'_\mathcal{X}
	=
	\prod_{n, \alpha>0} 
	(E^\alpha_n-\lambda_s\omega_\alpha) 
	(E^{-\alpha}_n+\lambda_s\omega_\alpha).
\end{align}
When the original HFB Hamiltonian $\mathcal{H}$ is time-reversal symmetric, 
from Eq.~(\ref{eq:pfH}) one obtains
\begin{equation}
	\mathrm{pf}\, \mathcal{H}_\mathcal{X}
	=
	\prod_{n, \alpha>0} (E^\alpha_n)^2.
\end{equation}
Substituting this into Eq.~(\ref{eq_m1}) together with $E^\alpha_n=E^{-\alpha}_n$, one sees
\begin{equation}
	\det \underline{\mathcal{W}}=+1.
\end{equation}
Substituting further this into Eq.~(\ref{eq_m2}), one sees
\begin{equation}
	\pf \mathcal{H}'_\mathcal{X}
	=
	\prod_{n, \alpha>0} 
	\left[ (E^\alpha_n)^2-(\lambda_s\omega_\alpha)^2 \right].
\end{equation}
Therefore from Eq.~(\ref{eq:formulaLE}) the number parity of the lowest-energy state under the constraint is given by
\begin{equation}
	\pi_{N}^\mathrm{L.E.} 
	=
	\prod_{n, \alpha>0} 
	\mathrm{sgn} \left[ (E^\alpha_n)^2-(\lambda_s\omega_\alpha)^2 \right].
\end{equation}
At $\lambda_s = 0$, $\pi_{N}^\mathrm{L.E.}=+1$, 
namely the lowest-energy state has even number parity. 
For $\lambda_s \ne 0$, 
the number parity changes according to the magnitude relation 
between $E^\alpha_n$ and $\lambda_s\omega_\alpha$. 
The number parity of the lowest-energy state thus changes according to the magnitude of $\lambda_s$.

\begin{figure}[t]
    \centering
    \includegraphics[width=0.8\columnwidth]{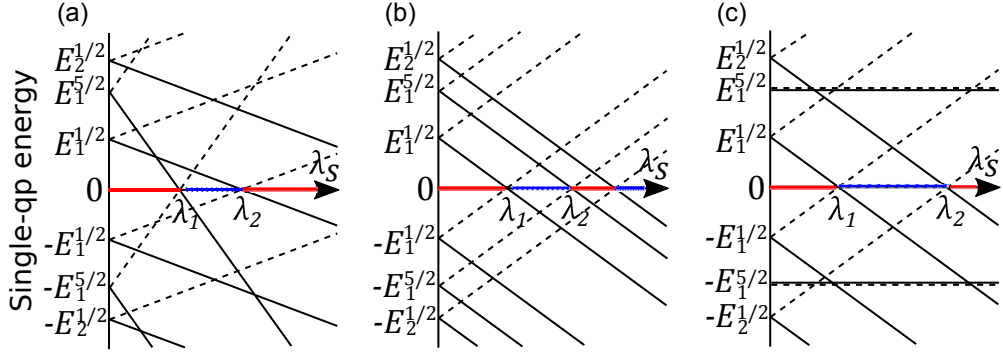}
    \caption{Schematic picture of the single-qp 
    levels with $\hat{J}_z$ eigenvalues $\Omega$ ($\Omega > 0$: solid line; $\Omega < 0$: dashed line) 
    as functions of the parameter $\lambda_s$ for the cases 
    employing the operator (a) $\hat{J}_z$,  (b) $\hat{J}_z/|J_z|$, and 
   (c) $\ket{J_z=1/2}\bra{J_z=1/2}-\ket{J_z=-1/2}\bra{J_z=-1/2}$ as the constraint operator $\hat{S}$.  
   Twelve levels with $\left|\Omega\right|=1/2^2, 5/2$ are shown. 
   At $\lambda_s=0$, the $n$th lowest (positive) single-qp energies with $\pm\Omega$ are degenerated and displayed as $E^{\Omega}_{n}$. 
   As increasing $\lambda_s$, the number parity of the lowest-energy state keeps $+1$ for $\lambda_s < \lambda_1$,  
   the sign changes at $\lambda_s = \lambda_1$, and the number parity is $-1$ for $\lambda_1<\lambda_s<\lambda_2$.
    \label{fig:method}}
\end{figure}

Figure~\ref{fig:method} shows a schematic picture of the single-qp energies under 
the constraint. 
When $\lambda_s$ is set to an appropriate value $\lambda_1 < \lambda_s < \lambda_2$, 
here a pair of levels intersect the axis at $\lambda_s=\lambda_1$, 
an odd number-parity state is automatically obtained as the lowest-energy state. 
Determined by the operator $\hat{S}$ are 
the orbital whose qp energy changes with an increase in $\lambda_s$ 
and the level which intersects the axis first. 
In this sense $\hat{S}$ is considered as a selector of the vacuum.
Note that when two levels cut across the axis by increasing $\lambda_s$, 
the number parity of the system becomes even corresponding to the two-qp excitation state. 
Specifically, when the $z$-component of the total angular momentum $\hat{J}_z$ 
is a symmetry of the intrinsic system, different vacuums are selected depending on the choice of $\hat{S}$. 
We here consider three cases. 
Case~(i); when $\hat{J}_z$ itself is taken as $\hat{S}$, 
each qp level rises or falls with the slope of the corresponding eigenvalue $J_z=\Omega$, see Fig.~\ref{fig:method}(a). 
As $\lambda_s$ increases, the orbital with high-$\Omega$ is preferably selected as the level to be excited. 
This is equivalent to the so-called non-collective cranking, 
where particles with high angular momentum about the symmetry axis are aligned. 
Case~(ii); let us look at the case of using $\hat{J}_z/|J_z|$ as $\hat{S}$. 
Since $\hat{J}_z/|J_z|$ is an operator 
that gives $+1$ for eigenstates with positive $\Omega$ and $-1$ for eigenstates with negative $\Omega$, 
each qp level separates out with the slope of $\pm 1$ independent of the magnitude of $\Omega$ 
with an increase in $\lambda_s$, see Fig.~\ref{fig:method}(b). 
Therefore, in this case, an orbital with a smaller qp energy is likely to be selected as the level to be excited. 
This corresponds to the two-Fermi level approach proposed in Ref.~\cite{ber09b}. 
As mentioned in Ref.~\cite{ber09b}, this choice may not work well for the case in which 
the single qp level density is high near the Fermi level and the spherical systems where we have $(2j+1)$-folded degeneracy. 
Introducing a kind of projection operator was conjectured in Ref.~\cite{ber09b} to resolve the issue 
and we realize the practical method as Case~(iii): Let us consider the case of using a time-odd projection operator 
$\ket{J_z=\Omega}\bra{J_z=\Omega}-\ket{J_z=-\Omega}\bra{J_z=-\Omega}$ as $\hat{S}$. 
This is a part of $\hat{J}_z/|J_z|$, 
and is an operator which works only for the state with a certain $\Omega$. 
Therefore, in this case, only the level carrying 
the specified eigenvalue $\pm\Omega$ splits for $\lambda_s \ne 0$, 
and the levels having other eigenvalues of $\hat{J}_z$ do not change even if $\lambda_s$ increases, see Fig.~\ref{fig:method}(c). 
Thus, one can select the state of interest easily.
The third choice is convenient for the practical use 
and we perform the calculations using this choice in the following investigation. 

We have generalized the method proposed in Ref.~\cite{ber09b} with a generic time-odd operator $\hat{S}$, 
and then proposed a practical choice for $\hat{S}$ of a time-odd projection operator. 
In Ref.~\cite{ber09b} the authors concluded that their approach had some difficulties compared with the conventional blocking method: 
Not all the qp states are easily accessible in their approach and it describes only the specific configurations; 
the method fails to describe a one-qp state when qp levels show degeneracies beyond the Kramers degeneracy;
the self-consistent calculations in a high qp-level density near the Fermi surface leads to numerical instabilities.
Our generalization and practical application of the method overcome these difficulties in some cases: 
Even when additional degeneracies are present beyond the Kramers degeneracy, our practical method gives a one-qp state with a desired quantum number as long as the $\hat{S}$ completely lifts those degeneracies, 
and the numerical instabilities can be less severe because the qp-level densities change as increasing $\lambda_s$ depending on the choice of $\hat{S}$. 
For instance, when $(2j+1)$-fold degeneracy is present in a spherical nuclei, one can selectively lower the single-particle energies of qp levels with a certain $\Omega$ to get the one-qp state with the quantum number. 
In neutron-drip line nuclei, however, a high density of state near the Fermi level causes the numerical instability as in the usual blocking method. 
Compared to the conventional blocking method, our method has advantage of being easy to code for calculating odd-mass nuclei: 
All we have to do is to add identity matrix with an appropriate coefficient to the HFB Hamiltonian as will be seen in detail in the next section, and then odd-mass nuclei are calculated as the lowest-energy solution of the HFB equation just like even--even nuclei.

\section{Numerical analysis for deformed neutron-rich Mg isotopes}\label{calculation}
\subsection{HFB equation for axially-symmetric nuclei with time-even and time-odd mean fields}
The coordinate-space HFB equation obtained by employing the local EDF containing time-even and time-odd parts reads~\cite{mat01, dob04}
\begin{equation}
\sum_{\sigma'=\pm\frac{1}{2}}
\begin{bmatrix} 
h^q_{\sigma \sigma'}(\boldsymbol{r})-\lambda^q \delta_{\sigma \sigma'}  & \tilde{h}^q_{\sigma \sigma'}(\boldsymbol{r})  \\ 
4\sigma \sigma' \tilde{h}^{q*}_{-\sigma -\sigma'}(\boldsymbol{r})  & -4\sigma\sigma' h^{q*}_{-\sigma -\sigma'}(\boldsymbol{r})+\lambda^q \delta_{\sigma \sigma'} 
\end{bmatrix} 
\begin{bmatrix} 
\varphi^q_{1,i}(\basissp) \\ 
\varphi^q_{2,i} (\basissp)  
\end{bmatrix}
= E^q_i 
\begin{bmatrix} 
\varphi^q_{1,i}(\basis) \\ 
\varphi^q_{2,i}(\basis) 
\end{bmatrix}, 
\label{HFB_eq}
\end{equation}
where $q$ stands for protons (p) and neutrons (n) 
in which the quasiparticles are assumed to be eigenstates of the third component of the isospin operator. 
The sp Hamiltonian $h$ consists of the mean-field (Kohn-Sham) potentials $\Gamma_t$ 
composed of the time-even and time-odd isoscalar ($t=0$) and isovector ($t=1$) densities as 
\begin{align}
&h^{\rm n}_{\sigma \sigma'} =
\left[ -\frac{\hbar^2}{2m} \bigtriangleup + \Gamma^\mathrm{even}_0 +\Gamma^\mathrm{odd}_0 +\Gamma^\mathrm{even}_1 +\Gamma^\mathrm{odd}_1 \right]_{\sigma \sigma'} , \\
&h^{\rm p}_{\sigma \sigma'} =
\left[-\frac{\hbar^2}{2m} \bigtriangleup + \Gamma^\mathrm{even}_0 +\Gamma^\mathrm{odd}_0 -\Gamma^\mathrm{even}_1 -\Gamma^\mathrm{odd}_1 +V_\mathrm{Coul}\right]_{\sigma \sigma'} .
\end{align}
Here, $V_\mathrm{Coul}$ is the Coulomb potential, and the explicit expressions of $\Gamma$ for the Skyrme-type EDF 
are shown in Appendix. 
Thanks to the time-reversal (anti-)symmetry of the potentials, one sees 
\begin{align}
\bar{h}^{{\rm n}}_{\sigma \sigma'}\coloneqq
&4\sigma\sigma' h^{{\rm n} *}_{-\sigma -\sigma'}= \left[-\frac{\hbar^2}{2m} \bigtriangleup + \Gamma^\mathrm{even}_0 -\Gamma^\mathrm{odd}_0 +\Gamma^\mathrm{even}_1 -\Gamma^\mathrm{odd}_1\right]_{\sigma \sigma'} , \\
\bar{h}^{{\rm p}}_{\sigma \sigma'}\coloneqq 
&4\sigma\sigma'h^{{\rm p}*}_{-\sigma -\sigma'}=\left[-\frac{\hbar^2}{2m} \bigtriangleup + \Gamma^\mathrm{even}_0 -\Gamma^\mathrm{odd}_0 -\Gamma^\mathrm{even}_1 +\Gamma^\mathrm{odd}_1 +V_\mathrm{Coul} \right]_{\sigma \sigma'}.
\end{align}
We employ the pairing EDF that contains only the time-even densities as described below, so that we see
\begin{equation}
4\sigma\sigma' \tilde{h}^{q*}_{-\sigma -\sigma'} =\tilde{h}^q_{\sigma \sigma'}.
\end{equation}

We solve the HFB equation (\ref{HFB_eq}) by assuming the axial and reflection symmetries so 
that the quasiparticles are labeled by $\{\Omega, \pi,q\}$, with $\Omega$ and $\pi$ being the 
$z$-component of the total angular momentum and parity, respectively.
In this case, the qp wave functions can be written in the form of 
\begin{equation}
\varphi^q_{a, n \Omega\pi}(\basis) =
\varphi^{q+}_{a, n \Omega\pi}(\varrho, z) \, \e^{\ii\Lambda^-\phi} \chi_{1/2}(\sigma) 
+ \varphi^{q-}_{a, n \Omega\pi}(\varrho, z) \, \e^{\ii\Lambda^+\phi} \chi_{-1/2}(\sigma), \quad (a=1, 2),
\label{wf_axial}
\end{equation}
where $\Lambda^{\pm}=\Omega\pm1/2$~\cite{vau73}, and 
$\varrho, z$, and $\phi$ are the cylindrical coordinates defining the three-dimensional position vector as $\boldsymbol{r}=(\varrho \cos \phi, \varrho \sin \phi, z)$, 
while $z$ is the chosen symmetry axis. And, the wave functions satisfy the following symmetry 
\begin{equation}
\varphi^{q \pm}_{a, n \Omega\pi}(\varrho, -z)=\pi (-1)^{\Lambda^\mp} \varphi^{q \pm}_{a, n \Omega\pi}(\varrho, z),  \qquad (a=1,2).
\end{equation}
The coordinate-space HFB equation has been solved under the assumption of the axial symmetry in many cases, however 
the time-reversal symmetry is often imposed~\cite{ter03,bla05,yos08a,oba08,pei08}. 
To keep the present paper self-contained, 
the mean-field potentials containing both the time-even and time-odd densities and currents 
in the cylindrical-coordinate representation are shown in Appendix. 
With the axial symmetry, the $\phi$ dependences of the sp and pair Hamiltonians are given by
\begin{equation}
h^q(\boldsymbol{r})=
\begin{bmatrix} 
h^q_{\ua\ua}(\varrho, z; l_z)& \e^{-\ii\phi}h^q_{\ua\da}(\varrho, z; l_z) \\ 
\e^{\ii\phi}h^q_{\da\ua}(\varrho, z; l_z) & h^q_{\da\da}(\varrho, z; l_z) 
\end{bmatrix},
\quad
\bar{h}^q(\boldsymbol{r})=
\begin{bmatrix} 
\bar{h}^q_{\ua\ua}(\varrho, z; l_z)& \e^{-\ii\phi}\bar{h}^q_{\ua\da}(\varrho, z; l_z) \\ 
\e^{\ii\phi}\bar{h}^q_{\da\ua}(\varrho, z; l_z) & \bar{h}^q_{\da\da}(\varrho, z; l_z) 
\end{bmatrix},
\end{equation}
and
\begin{equation}
\tilde{h}^q(\boldsymbol{r})=
\begin{bmatrix} 
\tilde{h}^q_{\ua\ua}(\varrho, z; l_z)& \e^{-\ii\phi}\tilde{h}^q_{\ua\da}(\varrho, z; l_z) \\ 
\e^{\ii\phi}\tilde{h}^q_{\da\ua}(\varrho, z; l_z) & \tilde{h}^q_{\da\da}(\varrho, z; l_z) 
\end{bmatrix},
\end{equation}
where $l_z = \frac{\del_\phi}{\ii}$, 
and thus the HFB equation in the $(\varrho, z)$ space for each $\{\Omega, \pi,q\}$ reads
\begin{equation}
\begin{pmatrix}
h^{\Omega \pi q}_{\ua\ua} -\lambda^q & h^{\Omega \pi q}_{\ua\da} & \tilde{h}^{\Omega \pi q}_{\ua\ua} & \tilde{h}^{\Omega \pi q}_{\ua\da} \\
h^{\Omega \pi q}_{\da\ua} & h^{\Omega \pi q}_{\da\da} -\lambda^q & \tilde{h}^{\Omega \pi q}_{\da\ua} & \tilde{h}^{\Omega \pi q}_{\da\da} \\
\tilde{h}^{\Omega \pi q}_{\ua\ua} & \tilde{h}^{\Omega \pi q}_{\ua\da} & -\bar{h}^{\Omega \pi q}_{\ua\ua} +\lambda^q & -\bar{h}^{\Omega \pi q}_{\ua\da} \\
\tilde{h}^{\Omega \pi q}_{\da\ua} & \tilde{h}^{\Omega \pi q}_{\da\da} & -\bar{h}^{\Omega \pi q}_{\da\ua} & -\bar{h}^{\Omega \pi q}_{\da\da} +\lambda^q
\end{pmatrix}
\begin{pmatrix}
\varphi^{q+}_{1,n \Omega \pi} \\ \varphi^{q-}_{1,n \Omega \pi} \\ \varphi^{q+}_{2,n \Omega \pi} \\ \varphi^{q-}_{2,n \Omega \pi}
\end{pmatrix}
=E^q_{n \Omega \pi}
\begin{pmatrix}
\varphi^{q+}_{1,n \Omega \pi} \\ \varphi^{q-}_{1,n \Omega \pi} \\ \varphi^{q+}_{2,n \Omega \pi} \\ \varphi^{q-}_{2,n \Omega \pi}
\end{pmatrix},
\label{2dHFB_eq}
\end{equation}
where $h^{\Omega \pi q}_{ss'}$, $\bar{h}^{\Omega \pi q}_{ss'}$, and $\tilde{h}^{\Omega \pi q}_{ss'}$ 
are defined by using $\Lambda_\ua \coloneqq \Lambda^-$ and $\Lambda_\da \coloneqq \Lambda^+$ as $h^{\Omega \pi q}_{ss'}(\varrho,z) \coloneqq h^{q}_{ss'}(\varrho,z; l_z=\Lambda_{s'})$, $\bar{h}^{\Omega \pi q}_{ss'}(\varrho,z) \coloneqq \bar{h}^{q}_{ss'}(\varrho,z; l_z=\Lambda_{s'})$, and $\tilde{h}^{\Omega \pi q}_{ss'}(\varrho,z) \coloneqq \tilde{h}^{q}_{ss'}(\varrho,z; l_z=\Lambda_{s'})$.

To describe odd-$A$ isotopes, we employ the time-odd projection operator to the states with a specific $\{\Omega, \pi, q\}$ quantum number, $\ket{\Omega\, \pi\, q}\bra{\Omega\, \pi\, q}-\ket{-\Omega\, \pi\, q}\bra{-\Omega\, \pi\, q}$, as the constraint operator $\hat{S}$. In other words, we introduce the Lagrange multiplier $\lambda_s$ for the specified $\{\Omega, \pi, q\}$ sector of the HFB equation (\ref{2dHFB_eq}) as
\begin{align}
\begin{pmatrix}
h^{\Omega \pi q}_{\ua\ua} -\lambda^q-\lambda_s & h^{\Omega \pi q}_{\ua\da} & \tilde{h}^{\Omega \pi q}_{\ua\ua} & \tilde{h}^{\Omega \pi q}_{\ua\da} \\
h^{\Omega \pi q}_{\da\ua} & h^{\Omega \pi q}_{\da\da} -\lambda^q-\lambda_s & \tilde{h}^{\Omega \pi q}_{\da\ua} & \tilde{h}^{\Omega \pi q}_{\da\da} \\
\tilde{h}^{\Omega \pi q}_{\ua\ua} & \tilde{h}^{\Omega \pi q}_{\ua\da} & -\bar{h}^{\Omega \pi q}_{\ua\ua} +\lambda^q-\lambda_s & -\bar{h}^{\Omega \pi q}_{\ua\da} \\
\tilde{h}^{\Omega \pi q}_{\da\ua} & \tilde{h}^{\Omega \pi q}_{\da\da} & -\bar{h}^{\Omega \pi q}_{\da\ua} & -\bar{h}^{\Omega \pi q}_{\da\da} +\lambda^q-\lambda_s
\end{pmatrix}
&\begin{pmatrix}
\varphi^{q+}_{1,n \Omega \pi} \\ \varphi^{q-}_{1,n \Omega \pi} \\ \varphi^{q+}_{2,n \Omega \pi} \\ \varphi^{q-}_{2,n \Omega \pi}
\end{pmatrix} \notag \\
=\left(E^q_{n \Omega \pi} - \lambda_s\right)
&\begin{pmatrix}
\varphi^{q+}_{1,n \Omega \pi} \\ \varphi^{q-}_{1,n \Omega \pi} \\ \varphi^{q+}_{2,n \Omega \pi} \\ \varphi^{q-}_{2,n \Omega \pi}
\end{pmatrix}.
\end{align}
Here, the chemical potential for protons or neutrons, $\lambda^q$, is adjusted so that the average particle number has the desired value, i.e. an odd number for odd-$A$ nuclei. 
We call below introducing $\lambda_s$ ``blocking" since this procedure is 
equivalent to the traditional blocking method.

\subsection{Numerical procedures}

We solve Eq.~(\ref{2dHFB_eq}) by diagonalizing the HFB Hamiltonian in the cylindrical-coordinate representation 
with the box boundary condition. 
We discretize the coordinates 
by $\varrho_i=(i-1/2)\times h$ $(i=1,2,\cdots, N_\rho)$ and $z_j=(j-1)\times h$ $(j=1,2,\cdots, N_z)$, with 
a lattice mesh size $h=$ 0.8 fm, and use 30 points for $N_\rho$ and $N_z$. 
Consequently, a qp wave function is expressed as a vector whose dimension is $N=4N_\rho N_z=3600$, and the HFB Hamiltonian is a matrix of size $N \times N$. 
The qp states are truncated according to the qp energy cutoff at 60 MeV, 
and the qp states up to $\Omega= 15/2$ 
with positive and negative parities are included.
The differential operators are represented by use of the 13-point formula of finite difference method.
For diagonalization of the Hamiltonian or Routhian, 
we use the LAPACK {\sc dsyevx} subroutine~\cite{LAPACK}. 
A modified Broyden's method~\cite{bar08} is utilized to calculate new densities during the self-consistent iteration. 
We consider that the calculation gets converged when both the difference of the total energy (in MeV) between two consecutive iterations and the deviations of the calculated particle numbers from the desired neutron and proton numbers become smaller than $10^{-4}$.
The Lagrange multiplier $\lambda_s$ is adjusted to $\lambda_s=(E^q_{1 \Omega \pi}+E^q_{2 \Omega \pi})/2$, 
where $E^q_{1 \Omega \pi}$ and $E^q_{2 \Omega \pi}$ are the lowest and the second lowest positive qp energies for the given $\{\Omega, \pi, q\}$ 
at each iteration so that only one pair of qp levels intersect the energy-zero axis. 

For the normal (particle-hole) part of a nuclear EDF, we employ the SLy4 functional~\cite{cha98}.  
The so-called na\"ive choice~\cite{sch10} is adopted for 
determining the coupling constants of the time-odd terms in the EDF except that the coupling constants of the terms of the form $\boldsymbol{s} \cdot \bigtriangleup \boldsymbol{s}$, where $\boldsymbol{s}$ is the spin density, are set to zero because the terms in some cases lead to divergences of the HFB iterative procedure~\cite{sch10}.
For the particle-particle channel, we adopt the density functional (\ref{E_pair}) which corresponds to the density-dependent contact interaction. 
The parameters are set as $V_1=1$ and $\gamma=1$ (surface pairing), and the pairing strength is taken as $V_0 = -430$ MeV fm$^3$ 
to reproduce approximately the experimental pairing gap of neutrons (1.28 MeV based on AME2016~\cite{wan17}) of $^{35}$Mg. 
The pairing gap is obtained by use of the three-point formula for the binding energy~\cite{sat98}, 
and the calculated $\Delta^{(3)}_{\mathrm{n}}$ for $^{35}$Mg is found to be 1.47 MeV. 

\subsection{Numerical results and discussion}\label{result}

To demonstrate the feasibility of our method, 
we perform the systematic calculation for the neutron-rich Mg isotopes with the mass number $A=34$ -- $40$. 
We exclude $^{32}$Mg and $^{33}$Mg in the present investigation, 
where the loss of spherical magic number of 20 has been under debate, 
because the shape fluctuation and the correlation beyond the mean-field approximation 
may be significant in $^{32}$Mg~\cite{kim02,rod02, hin11}, 
and many-particle many-hole states with different shape deformation 
may coexist in $^{33}$Mg~\cite{kim11,kim11b} as mentioned slightly below.

We tried blocking each of $\Omega^\pi=1/2^{\pm}, 3/2^{\pm}, 5/2^{\pm}$, and $7/2^{\pm}$ orbitals for odd-mass isotopes, and
the ground state was obtained by blocking 
the orbital with $\Omega^\pi=3/2^-, 5/2^-$, and $1/2^-$ 
in $^{35}$Mg, $^{37}$Mg, and $^{39}$Mg, respectively. 
The calculation may be in contradiction with the observation for $^{35}$Mg, 
where $J^\pi=3/2^-$, a head of the $K^{\pi}=1/2^-$ band, is suggested for the ground state~\cite{mom17}.  
We found that the binding energy obtained by blocking the $\Omega^\pi=1/2^-$ orbital 
is shallower by 1.0  MeV.
For $^{37}$Mg, the measurements suggest that 
the $\ell = 1$ component is dominant in the ground state~\cite{kob14,tak14}, while the 
$\Omega^\pi=5/2^-$ orbital contains angular momenta higher than $\ell = 3$.  
Below, we are going to discuss $^{37}$Mg on this point. 
It is noted here that the neutron superfluidity vanishes in $^{35,37}$Mg. 

\begin{figure}[t]
    \centering
    \includegraphics[width=0.4\columnwidth]{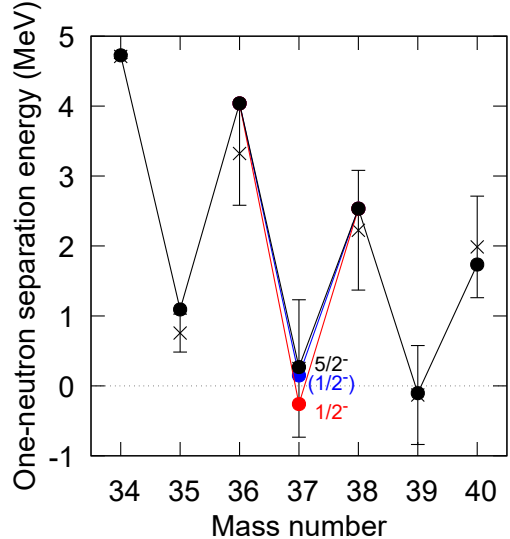}
    \caption{Calculated one-neutron separation energies $S_n$ of Mg isotopes denoted by closed circles 
    together with the experimental data from AME2016~\cite{wan17} denoted by crosses with error bars. 
    For $^{37}$Mg, the results obtained by blocking the 
    $\Omega^\pi=1/2^-$ orbital are also shown. 
    The symbols for $(1/2^-)$ indicate the results obtained by ignoring the time-odd mean fields. 
    \label{fig:Mg1}}
\end{figure}

Figure~\ref{fig:Mg1} shows the calculated one-neutron separation energies $S_{\mathrm{n}}$ compared 
with the experimental or evaluated data obtained from AME2016.  
Here, $S_n$ is calculated as
\begin{align}
	S_n \coloneqq -\left[ B(N,Z) - B(N-1,Z) \right],
\end{align}
where $B(N,Z)$ is the (negative) binding energy of the nucleus with $N$ neutrons and $Z$ protons. 
A nice agreement within the error range and the odd-even staggering feature in the binding energies can be seen. 
It is noted that 
we need the binding energy of $^{33}$Mg for the calculation of $S_{\mathrm{n}}$ of $^{34}$Mg. 
We obtained the near-degeneracy by blocking the $1/2^-$ and $7/2^-$ orbitals for $^{33}$Mg. 
For $^{39}$Mg, 
the calculated one-neutron separation energy is $S_n = -0.10$ MeV, 
the calculated chemical potential is $\lambda^{\textrm n}=-0.98$ MeV, and 
the qp energy of the blocked orbital is 0.83 MeV. 
Thus, the $^{39}$Mg nucleus is 
unstable with respect to the neutron emission, 
but bound, though quite loosely, in the sense that the sp energy of the last occupied neutron is negative 
in the present calculation. 
On the other hand, we could not find the bound-state solutions for $^{41,42}$Mg. 

\begin{figure}[t]
    \centering
    \includegraphics[width=0.45\columnwidth]{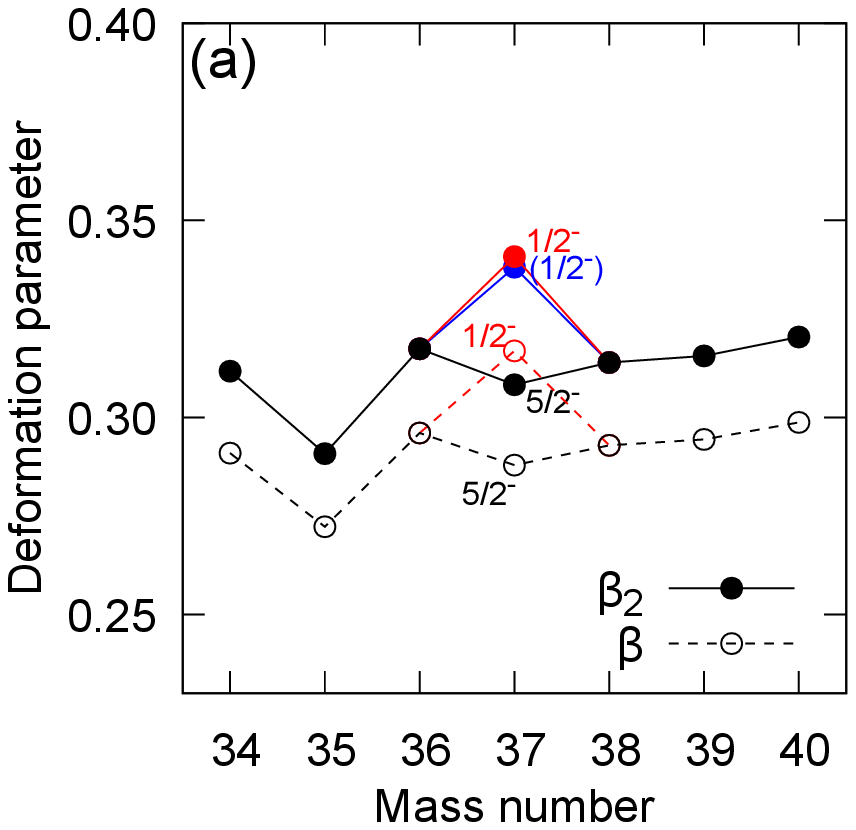}
    \hspace{10pt}
    \includegraphics[width=0.45\columnwidth]{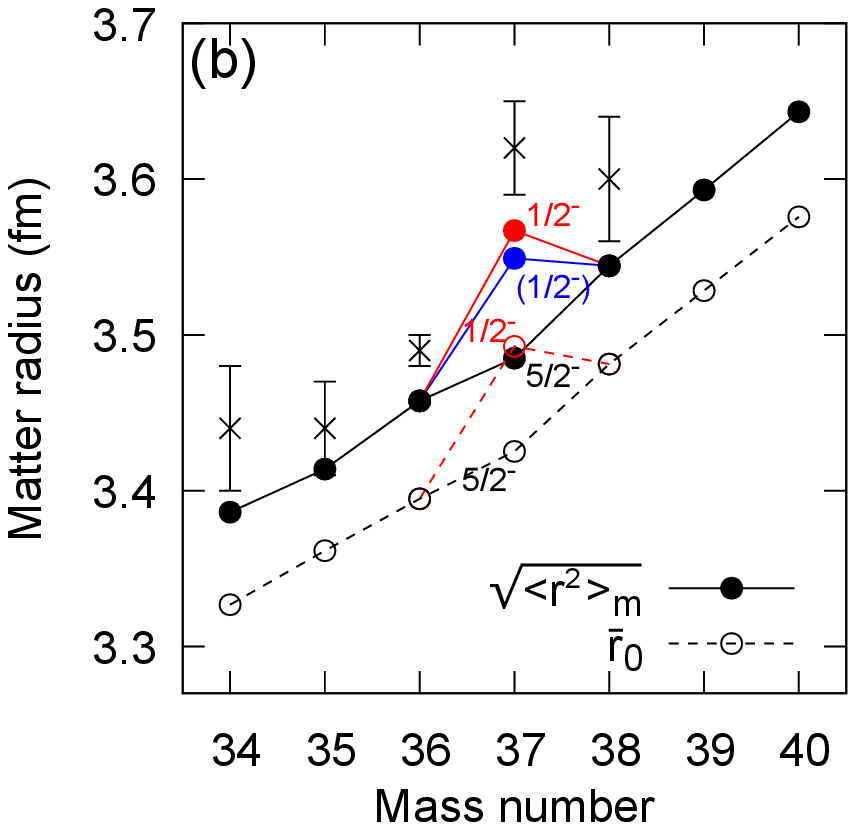}
    \caption{(a) Calculated deformation parameters of Mg isotopes. 
    The closed circles denote the quadrupole-deformation parameters $\beta_2$. 
    For reference, the calculated $\beta$ values defined in Eq.~(\ref{eq:beta_r0}) are plotted by the open circles connected by the dashed line. 
    (b) Matter radii of Mg isotopes. 
    The closed circles denote the calculated matter radii $\sqrt{\braket{r^2}_m}$, compared with the observation denoted by the crosses with error bars taken from Ref.~\cite{wat14}. 
    For reference, the calculated $\bar{r}_0$ values defined in Eq.~(\ref{eq:beta_r0}) are plotted by the open circles connected by the dashed line. 
    As in Fig.~\ref{fig:Mg1}, the results obtained by blocking the 
    $\Omega^\pi=1/2^-$ orbital are also shown for $^{37}$Mg. 
    The symbols for $(1/2^-)$ indicate the results obtained by ignoring the time-odd mean fields. 
    \label{fig:Mg2}}
\end{figure}
We show in Fig.~\ref{fig:Mg2} the calculated quadrupole-deformation parameters $\beta_2$ 
and matter radii $\sqrt{\langle r^2\rangle_{\mathrm{m}}}$, which are defined by 
\begin{align}
	&\beta_2 \coloneqq \dfrac{4\pi}{5 A \langle r^2\rangle_{\mathrm{m}}} \int d\boldsymbol{r} r^2Y_{20}(\hat{r})\rho_0(\boldsymbol{r}),\\
	&\sqrt{\langle r^2\rangle_{\mathrm{m}}} \coloneqq \sqrt{\frac{1}{A} \int d\boldsymbol{r} r^2 \rho_0(\boldsymbol{r})},
\end{align}
with closed circles. 
As shown in Fig.~\ref{fig:Mg2}(a), the Mg isotopes under study are calculated to be constantly deformed 
and this is consistent with the preceding theoretical predictions 
on the Mg isotopes near the drip line~\cite{ter97, rod02,yos09,li12,shi16,nak18,now09,cau14}. 
The odd-even staggering in deformation 
is faint compared with that of the one-neutron separation energies.
Therefore, 
the odd-even staggering seen in the binding energy of these isotopes 
can be attributed to the pair correlation mainly. 
Let us discuss the systematic feature in matter radii. 
Figure~\ref{fig:Mg2}(b) shows the calculated matter radii 
$\sqrt{\langle r^2\rangle_{\mathrm{m}}}$ compared with the observation based on the 
reaction cross section measurement~\cite{wat14}.  
Except for $^{37}$Mg, the present calculation reproduces the isotopic dependence observed experimentally. 
However, we see a systematic underestimation. 
This is mainly because the calculation gives a systematic over binding. 
The irregular dependence revealed in $^{34\textrm{--}36}$Mg by the reaction cross section measurement~\cite{tak14} 
is well described; the matter radius of $^{35}$Mg is smaller than the average of the radii of the neighboring isotopes of $^{34,36}$Mg. 
The suppression of the matter radius in $^{35}$Mg would be attributed to the smaller deformation than in the neighboring isotopes as shown in Fig.~\ref{fig:Mg2}(a). 
To make this point clear, we shall adopt the following alternative expressions of the deformation and matter radius for axially-symmetric nuclei~\cite{wat14, nak18}: 
\begin{align}
	\langle x^2 \rangle_{\mathrm{m}} 
	= 
	\langle y^2 \rangle_{\mathrm{m}} 
	=
	\frac{1}{A} \int d\boldsymbol{r} \frac{\varrho^2}{2} \rho_0(\boldsymbol{r})
	&\eqqcolon
	\frac{\bar{r}_0^2}{3} \exp \left(-\sqrt{\frac{5}{4\pi}}\beta \right), \notag \\
	\langle z^2 \rangle_{\mathrm{m}} 
	= 
	\frac{1}{A} \int d\boldsymbol{r} z^2 \rho_0(\boldsymbol{r})
	&\eqqcolon
	\frac{\bar{r}_0^2}{3} \exp \left(2 \sqrt{\frac{5}{4\pi}}\beta \right).
	\label{eq:beta_r0}
\end{align}	
One can see the relation as
\begin{align*}
	&\beta
	=
	\frac{1}{3}
	\sqrt{\frac{4\pi}{5}}
	\ln
	\left(
	\frac{1+2\sqrt{\frac{5}{4\pi}}\beta_2}
	{1-\sqrt{\frac{5}{4\pi}}\beta_2}
	\right)
	=\beta_2 -\frac{1}{4}\sqrt{\frac{5}{\pi}}\beta_2^2 +\mathcal{O}(\beta_2^3),  \\
	&\bar{r}_0
	=
	\sqrt{\langle r^2\rangle_{\mathrm{m}}}
	\left[
	\left(1-\sqrt{\frac{5}{4\pi}}\beta_2\right)^2
	\left(1+2\sqrt{\frac{5}{4\pi}}\beta_2\right)
	\right]^{\frac{1}{6}}
	=
	\sqrt{\langle r^2\rangle_{\mathrm{m}}}
	\left[
	1-\frac{5}{8\pi}\beta_2^2+\mathcal{O}(\beta_2^3)
	\right], 
\end{align*}
and $\beta \approx \beta_2$ and $\bar{r}_0 \approx \sqrt{\langle r^2\rangle_{\mathrm{m}}}$ for small deformation. 
The definitions of the deformation parameter $\beta$ and radius $\bar{r}_0$ 
in Eq.~(\ref{eq:beta_r0}) 
guarantee the volume conservation 
of a spheroidal nucleus, whose density is given by $\rho_0 = A/V$ if $\varrho^2/a^2 + z^2/b^2 \le 1$ and zero if $\varrho^2/a^2 + z^2/b^2 > 1$ with $V=4\pi a^2 b /3 = 4\pi/3 (5\bar{r}_0/3)^{3/2}$~\cite{hil53}. 
In terms of the parameters $\beta$ and $\bar{r}_0$, the effects of the deformation and the spatial extension are decoupled in the spheroidal nuclei with a sharp surface owing to the volume conservation.
We show the calculated $\beta$ and $\bar{r}_0$ values in Fig.~\ref{fig:Mg2} with the open circles.
The suppression of the matter radius $\sqrt{\langle r^2\rangle_{\mathrm{m}}}$ in $^{35}$Mg disappears in $\bar{r}_0$ while the deformation is reduced in  $^{35}$Mg compared with $^{34,36}$Mg in terms of $\beta$ as well as $\beta_2$. This clearly shows that the reduction of the deformation accounts for the suppression of the matter radius in $^{35}$Mg.

The discrepancy in the matter radius between the calculation and the observation for $^{37}$Mg is 
due to the suppression of spatial extension caused by 
the high centrifugal barrier of the blocked orbital of $[312]5/2$ stemming from the $f_{7/2}$ shell. 
As mentioned above, the experimental measurements suggest that 
the ground state in $^{37}$Mg is dominated by the $p$-wave~\cite{kob14,tak14}. 
When the deformation develops further, the $[312]5/2$ orbital crosses with the $[321]1/2$ orbital. 
The latter orbital is originating from the $p_{3/2}$ shell. 
Therefore, blocking the $\Omega^\pi=1/2^-$ orbital is worth investigating. 
Indeed, the deformed halo structure in $^{37}$Mg has been studied by 
assuming a high deformation of $\beta_2 \sim 0.5$ in a deformed Woods-Saxon potential 
to put a neutron in the [321]1/2 orbital on the even--even $^{36}$Mg nucleus~\cite{ura17}. 
We show in Figs.~\ref{fig:Mg1} and \ref{fig:Mg2} the results obtained by blocking the $\Omega^\pi=1/2^-$ orbital for $^{37}$Mg 
as indicated by red circles. 
In Fig.~\ref{fig:Mg2}(a), the matter radius increases by 0.08 fm by blocking the $\Omega^\pi=1/2^-$ orbital.  
Then, we can see a sudden enhancement from $^{36}$Mg. 
As expected, 
the deformation develops that we can see in Fig.~\ref{fig:Mg2}(a). 
The matter quadrupole deformation obtained is $\beta_2 =  0.35$, 
and we see an increase in $\beta_2$ by 0.03, 
which is far lower than the phenomenological value~\cite{ura17}. 
Note that $\beta$ and $\bar{r}_0$ contribute cooperatively to the enhancement of $\sqrt{\langle r^2\rangle_{\mathrm{m}}}$ in $^{37}$Mg  indicating dilution of the density with deformation, in contrast to the case of $^{35}$Mg, where only the deformation contributes to the suppression of the matter radius as seen above and in Ref.~\cite{nak18}.
The total binding energy calculated by blocking the $\Omega^\pi=1/2^-$ orbital is shallower by 0.53 MeV, 
which we see in Fig.~\ref{fig:Mg1}. 
We found that the neutrons in $^{37}$Mg obtained by blocking the $\Omega^\pi=1/2^-$ orbital 
are paired. 
The calculated chemical potential and the qp energy of the blocked orbital is 
$-2.70$ MeV and $2.20$ MeV, respectively, while 
the sp energy of the last occupied neutron is $-1.1$ MeV obtained by blocking the $\Omega^\pi=5/2^-$ orbital. 
The asymptotic behavior of the last occupied orbital is given as 
$r\varphi_{2,i}(\boldsymbol{r})\sim \exp[-\sqrt{-2m(\lambda+E_i)}r/\hbar]$ 
for a paired system and $\sim \exp[-\sqrt{-2m \epsilon_i}r/\hbar]$ for a unpaired system with $\epsilon_i$ being the sp energy~\cite{dob84}. 
Thus, we have enhancement in the radius by blocking the $\Omega^\pi=1/2^-$ orbital though 
the chemical potential is not very shallow. 
This can be considered as unpaired-particle haloing~\cite{nak18}.
It should be noted that in Ref.~\cite{nak18} the self-consistent HFB calculation with a semirealistic interaction produces nicely the enhancement of the matter radius for the ground state in $^{37}$Mg. 
Our framework is similar to that of Ref.~\cite{nak18}. 
The interactions used are different. 
This suggests that the ground-state property in $^{37}$Mg can impose restriction on the EDF.

\begin{figure}[t]
    \centering
    \includegraphics[width=0.45\columnwidth]{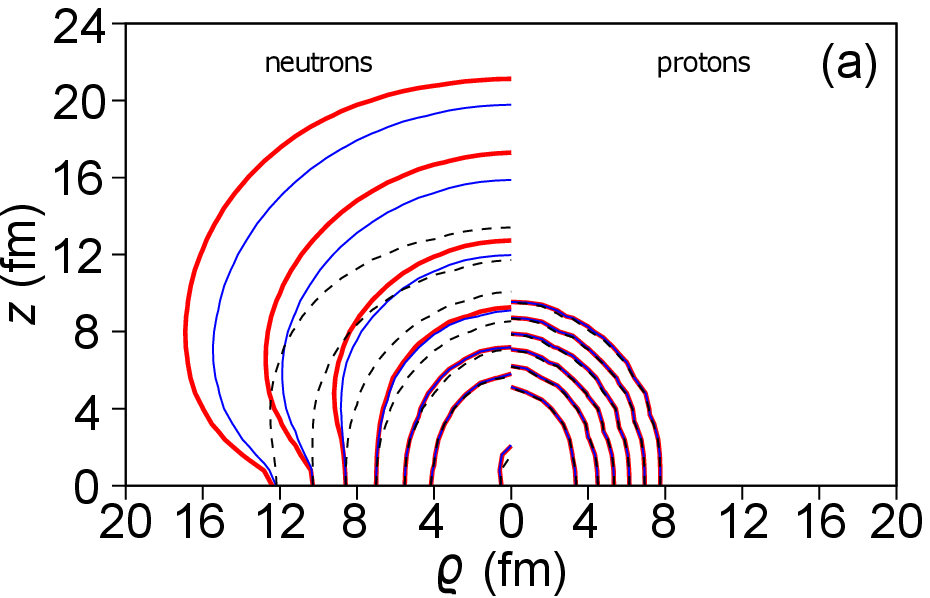}
    \hspace{10pt}
    \includegraphics[width=0.45\columnwidth]{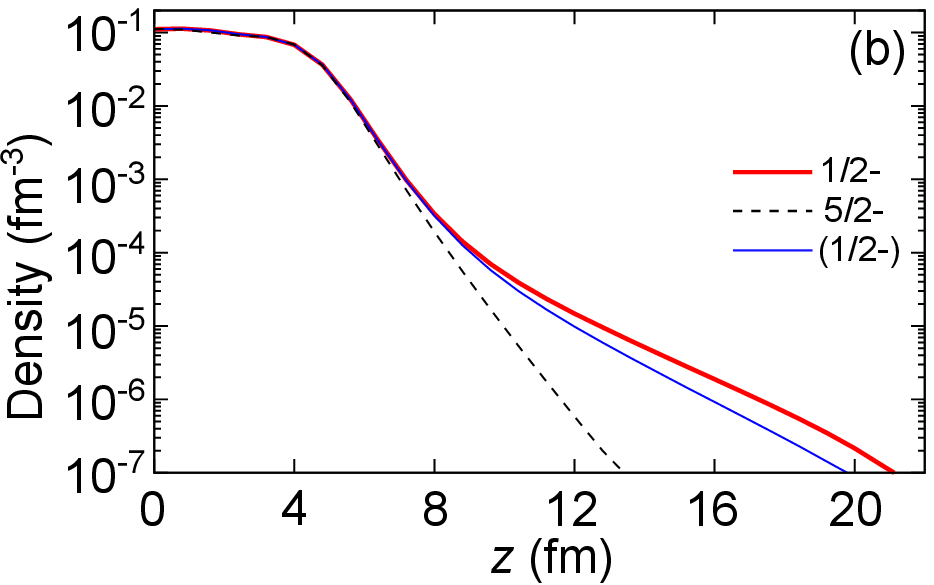}
    \caption{(a) Contour plot of neutron and proton densities on the $\varrho$-$z$ plane for $^{37}$Mg.  
    Positions of the density 
    with 0.1, 0.01, $10^{-3}$, $10^{-4}$, $10^{-5}$, $10^{-6}$, and $10^{-7}$ fm$^{-3}$ are presented. 
    (b) Neutron density distributions along the symmetry axis. 
    The results obtained by blocking the $\Omega^\pi=1/2^-$ and $5/2^-$ orbitals are shown. 
    In the case of $\Omega^\pi=1/2^-$, shown are also the result obtained by ignoring the time-odd mean fields.
    \label{fig:Mg37}}
\end{figure}

To see the spatial structure of $^{37}$Mg, 
we draw the calculated density distributions in terms of the equidensity lines on the 
$\varrho$-$z$ plane in Fig.~\ref{fig:Mg37}(a). 
The contour lines are depicted in a logarithmic scale at 0.1 fm$^{-3}$ 
down to $10^{-7}$ fm$^{-3}$. 
In Fig.~\ref{fig:Mg37}(b), shown are the density distributions of neutrons 
along the symmetry axis (at $\varrho = 0.4$ fm) also in a logarithmic scale. 
The density distributions obtained by blocking the $\Omega^\pi=5/2^-$ and $1/2^-$ orbitals are presented. 
In the case of $\Omega^\pi=5/2^-$, the density distribution of neutrons is well localized in the center 
though the spatial extension is visible compared with that of protons, forming the neutron skin.
Blocking the $\Omega^\pi=1/2^-$ orbital changes drastically the distribution of neutrons. 
A long tail emerges, interpreted as the neutron halo. 
The spatial distribution is extended toward the symmetry axis, forming a peanut shape. 
This is consistent with the previous calculation~\cite{nak18}, 
and results from the $p$-wave dominance near the continuum threshold~\cite{mis97,ham04,yos05,zho10}. 

The time-reversal symmetry is intrinsically broken in 
the odd-mass isotopes possessing nonzero spin, 
so that the time-odd components in the mean field may be activated. 
Let us discuss finally the roles of the time-odd mean fields in $^{37}$Mg by blocking the 
$\Omega^\pi=1/2^-$ orbital. 
The total binding energy is affected by 0.41 MeV, as shown in Fig.~\ref{fig:Mg1}(a). 
Here the time-odd fields are set to zero, i.e. equivalent to the equal-filling approximation. 
This is only 0.16\% to the total binding energy, and is negligibly small. 
A tiny effect on the nuclear mass has been brought out by the systematic calculation~\cite{sch10}. 
Accordingly, the deformation property is hardly influenced by the time-odd fields as shown in Fig.~\ref{fig:Mg1}(b). 
The deformation parameter for neutrons is reduced by 0.01. 
The radius of neutrons thus calculated is lessened by 0.02 fm, while the protons are not affected. 
Then the matter radius is reduced by 0.02 fm as shown in Fig.~\ref{fig:Mg2}. 
The chemical potential and the qp energy of the blocked orbital is $-2.73$ MeV and $1.92$ MeV, 
respectively. 
Then, the qp energy of the last neutron is lowered by 0.28 MeV by ignoring the time-odd fields. 
It seems that this shift is also negligible. 
However, the asymptotic behavior of the halo structure is sensitive 
to the exponent of the qp wave function. 
Indeed, as shown in Fig.~\ref{fig:Mg37}, the tail structure is influenced by the time-odd fields. 
Whether the time-odd mean fields enhance or reduce the halo structure depends on 
the EDF employed. 
The spin-density term in the EDF, the first term of the right-hand side in Eq.~(\ref{eq:Hodd}), is responsible for the enhancement in the radius: 
The matter radius of $^{37}$Mg obtained by ignoring all the time-odd fields except those derived from the spin-density term has been no more than 0.02\% different from the result obtained by including all the time-odd fields. 
The reaction observables sensitive to the outer surface of the halo nucleus can 
put constraint on the time-odd coupling constants, especially on the coefficient of the spin-density term, of the Skyrme EDF that are uncertain. 

\section{Summary}\label{summary}

We have found relationships among 
the particle-number parity, 
the Bogoliubov transformation, 
and the time-reversal symmetry of the Hartree--Fock--Bogoliubov (HFB) Hamiltonian. 
Then we showed that 
the lowest-energy solution of the HFB equation has the even particle-number parity 
as long as the time-reversal symmetry is conserved in the HFB Hamiltonian without null eigenvalues. 
Based on this finding, we gave foundation of a method for solving the 
HFB equation to describe the ground state of odd-mass nuclei by 
employing an appropriate time-reversal anti-symmetric constraint operator to the Hamiltonian.   
With this procedure, 
one can obtain directly the ground state of an odd-mass nucleus 
as a self-consistent solution of the cranked-HFB-type equation,  
while the ground state of an odd-mass nucleus is described as a one-quasiparticle 
excitation of a neighboring even--even nucleus in a usual procedure. 
This method is further applicable to the low-lying two-quasiparticle excitations in even--even nuclei.   
As a numerical example, we applied this method to the neutron-rich Mg isotopes close 
to the drip line, and showed that the anomalous increase in the matter radius of $^{37}$Mg 
is well described when a neutron occupies the  
low-$\Omega$ orbital 
 in the framework of the nuclear energy-density functional method. 
We found that the time-odd mean fields have little influence on the total binding energy, 
but an appreciable impact on the asymptotic behavior of the halo structure.

\section*{Acknowledgment}
The authors thank A.~Afanasjev, M.~Grasso, N.~Itagaki, M.~Matsuo, T.~Nakatsukasa, and D.~Vretenar for 
valuable discussions and comments. 
This work was supported by the JSPS KAKENHI (Grants No. JP19K03824 and No. JP19K03872), 
and the JSPS-NSFC Bilateral Program for Joint Research Project on 
``Nuclear mass and life for unraveling mysteries of the r-process''. 
The numerical calculations were performed on CRAY XC40 at the Yukawa Institute for Theoretical Physics, Kyoto University.

\appendix

\section{Mean-field potentials for axially-symmetric nuclei}
In this appendix we shall give the explicit expressions of the mean-field potentials obtained from the Skyrme EDF and the pairing EDF for axially-symmetric nuclei.

\subsection{Mean-field potentials in the Skyrme EDF for axially-symmetric nuclei}
The Skyrme EDF consists of the time-even and time-odd parts~\cite{eng75}:
\begin{equation}
\mathcal{E}_\mathrm{Sky}= \int d\boldsymbol{r} \sum_{t=0,1} \left[\mathscr{H}_t^\mathrm{even}(\boldsymbol{r}) +\mathscr{H}_t^\mathrm{odd}(\boldsymbol{r}) \right],
\end{equation}
where 
\begin{align}
&\mathscr{H}_t^\mathrm{even} =C_t^\rho[\rho_0] \rho_t^2+ C_t^{\bigtriangleup \rho} \rho_t \bigtriangleup \rho_t +C_t^\tau \rho_t \tau_t +C_t^{J} \overleftrightarrow{J_t}^2 +C_t^{\nabla J} \rho_t \boldsymbol{\nabla}\cdot \boldsymbol{J}_t, \\
\label{eq:Hodd}
&\mathscr{H}_t^\mathrm{odd} =C_t^s[\rho_0] \boldsymbol{s}_t^2+ C_t^{\bigtriangleup s} \boldsymbol{s}_t \cdot \bigtriangleup \boldsymbol{s}_t +C_t^T \boldsymbol{s}_t \cdot \boldsymbol{T}_t +C_t^j \boldsymbol{j}_t^2 +C_t^{\nabla j} \boldsymbol{s}_t \cdot (\boldsymbol{\nabla} \times \boldsymbol{j}_t),
\end{align}
with $t=0$ and 1 denoting isoscalar and isovector, respectively. 
Here, the definitions of the densities and currents are given in Ref.~\cite{eng75}. 
Then the mean-field potentials are given by the functional derivatives as~\cite{eng75, dob95}
\begin{align}
&\Gamma^\mathrm{even}_{\mathrm{Sky}, t}(\boldsymbol{r})=-\boldsymbol{\nabla}\cdot M_t(\boldsymbol{r}) \boldsymbol{\nabla} +U_t(\boldsymbol{r}) +\frac{1}{2\ii} [ \overleftrightarrow{\nabla\sigma} \overleftrightarrow{B}_t(\boldsymbol{r}) +\overleftrightarrow{B}_t(\boldsymbol{r}) \overleftrightarrow{\nabla\sigma} ] +\delta_{0t} U_0'(\boldsymbol{r}) \label{Gammaeven}, \\
&\Gamma^\mathrm{odd}_{\mathrm{Sky}, t}(\boldsymbol{r})=-\boldsymbol{\nabla}\cdot (\boldsymbol{\sigma}\cdot \boldsymbol{C}_t(\boldsymbol{r})) \boldsymbol{\nabla} +\boldsymbol{\sigma}\cdot \boldsymbol{\Sigma}_t(\boldsymbol{r}) +\frac{1}{2\ii} [ \boldsymbol{\nabla}\cdot \boldsymbol{I}_t(\boldsymbol{r}) +\boldsymbol{I}_t(\boldsymbol{r})\cdot \boldsymbol{\nabla} ] \label{Gammaodd}.
\end{align}
Here we used the symbol $\overleftrightarrow{\nabla\sigma}\coloneqq\boldsymbol{\nabla} \otimes \boldsymbol{\sigma}$, 
and, 
\begin{align}
&U_t=2C_t^\rho \rho_t +2C_t^{\bigtriangleup\rho}\bigtriangleup\rho_t+C^\tau_t\tau_t+C^{\nabla J}_t \boldsymbol{\nabla}\cdot \boldsymbol{J}_t, \\
&\boldsymbol{\Sigma}_t=2C^s_t \boldsymbol{s}_t +2C^{\bigtriangleup s}\bigtriangleup\boldsymbol{s}_t+C^T_t\boldsymbol{T}_t+C^{\nabla j}_t \boldsymbol{\nabla}\times \boldsymbol{j}_t , \\
&M_t=C^\tau_t \rho_t , \\
&\boldsymbol{C}_t=C^T_t\boldsymbol{s}_t , \\
&\overleftrightarrow{B}_t=2C^J_t \overleftrightarrow{J}_t-C^{\nabla J} \overleftrightarrow{\nabla}\rho_t , \\
&\boldsymbol{I}_t=2C^j_t\boldsymbol{j}_t+C^{\nabla j} \boldsymbol{\nabla} \times \boldsymbol{s}_t , \\
&U_0'=\sum_{t=0,1} \left(\frac{\del C^\rho_t}{\del \rho_0} \rho_t^2+\frac{\del C^s_t}{\del \rho_0} \boldsymbol{s}_t^2 \right),
\end{align}
where $\nabla_{\mu\nu}\coloneqq\sum_{\kappa} \epsilon_{\mu\nu\kappa}\nabla_\kappa$.

The mean-field potentials $\Gamma_{\mathrm{Sky}}$ are composed of the densities and currents, 
so let us show the expressions in the cylindrical coordinates $\boldsymbol{r}=(\varrho, \phi, z)$ employing the ans\"atze (\ref{wf_axial}). 
The time-even densities are given as
\begin{align}
&\rho^q =\sum_i \left[(\vap)^2+(\vam)^2 \right] ,\\
&\left\{\begin{aligned}
&J_{\varrho\phi}^q=-\sum_i (\vap\del_\varrho\vam-\vam\del_\varrho\vap) \\
&J_{\phi\varrho}^q=\sum_i \frac{\Lambda^+_i+\Lambda^-_i}{\varrho} \vap \vam \\
&J_{\phi z}^q =\sum_i \left[\frac{\Lambda^-_i}{\varrho}(\vap)^2-\frac{\Lambda^+_i}{\varrho}(\vam)^2\right]\\
&J_{z\phi}^q =-\sum_i (\vap\del_z\vam -\vam\del_z\vap) \\
&J_{\varrho\varrho}^q=J_{\phi\phi}^q=J_{zz}^q=J_{\varrho z}^q=J_{z\varrho}^q=0
\end{aligned}\right. ,
\end{align}
and 
\begin{align}
&\tau^q=\sum_i \left[ (\del_\varrho\vap)^2+(\del_\varrho \vam)^2 +\left(\frac{\Lambda^-_i}{\varrho}\vap\right)^2 +\left(\frac{\Lambda^+_i}{\varrho}\vam\right)^2+(\del_z\vap)^2+(\del_z\vam)^2\right].
\end{align}
The time-odd densities are 
\begin{align}
&\left\{\begin{aligned}
&s_\varrho^q=-2 \sum_i \vap \vam \\
&s_z^q=-\sum_i\left[(\vap)^2-(\vam)^2\right] \\
&s_\phi^q=0 
\end{aligned} \right. , \\
&\left\{\begin{aligned}
&j_\phi^q=-\sum_{i}\left[\frac{\Lambda^-_i}{\varrho} (\vap)^2 +\frac{\Lambda^+_i}{\varrho}(\vam)^2\right] \\
&j_\varrho^q=j_\phi^q=0
\end{aligned}\right. ,
\end{align}
and 
\begin{align}
&\left\{\begin{aligned}
&T_\varrho^q=-2\sum_i \left[ (\del_\varrho \vap)(\del_\varrho\vam)+(\del_z\vap)(\del_z \vam)+\frac{\Lambda^-_i\Lambda^+_i}{\varrho^2}\vap\vam\right] \\
&T_z^q=-\sum_i \left[(\del_\varrho\vap)^2-(\del_\varrho \vam)^2 +\left(\frac{\Lambda^-_i}{\varrho}\vap\right)^2-\left(\frac{\Lambda^+_i}{\varrho}\vam\right)^2+(\del_z\vap)^2-(\del_z \vam)^2 \right] \\
&T_\phi^q=0
\end{aligned} \right. .
\end{align}

Substituting these densities into Eqs.~(\ref{Gammaeven}) and (\ref{Gammaodd}), one obtains 
\begin{align}
\Gamma^\mathrm{even}_{\mathrm{Sky}, t}(\boldsymbol{r})&=
\begin{bmatrix} 
\Gamma^\mathrm{even}_{\mathrm{Sky}, t\, \ua\ua}(\varrho, z; l_z)& \e^{-\ii\phi}\Gamma^\mathrm{even}_{\mathrm{Sky}, t\, \ua\da}(\varrho, z; l_z) \\ 
\e^{\ii\phi}\Gamma^\mathrm{even}_{\mathrm{Sky}, t\, \da\ua}(\varrho, z; l_z) & \Gamma^\mathrm{even}_{\mathrm{Sky}, t\, \da\da}(\varrho, z; l_z) 
\end{bmatrix}, \\
\Gamma^\mathrm{odd}_{\mathrm{Sky}, t}(\boldsymbol{r})&=
\begin{bmatrix} \Gamma^\mathrm{odd}_{\mathrm{Sky}, t\, \ua\ua}(\varrho, z; l_z) & \e^{-\ii\phi} \Gamma^\mathrm{odd}_{\mathrm{Sky}, t\, \ua\da}(\varrho, z; l_z) \\ 
\e^{\ii\phi}\Gamma^\mathrm{odd}_{\mathrm{Sky}, t\, \da\ua}(\varrho, z; l_z) & \Gamma^\mathrm{odd}_{\mathrm{Sky}, t\, \da\da}(\varrho, z; l_z) 
\end{bmatrix},
\end{align}
where 
\begin{align}
\Gamma^\mathrm{even}_{\mathrm{Sky}, t\, \ua\ua}&=
-(\del_\varrho M_t) \del_\varrho -(\del_z M_t) \del_z -M_t\bigtriangleup +U_t +\delta_{0t} U_0'+ B_{t\, \phi z}\frac{l_z}{\varrho} \\
\Gamma^\mathrm{even}_{\mathrm{Sky}, t\, \ua\da}&=
-K_t- B_{t\, \varrho\phi} \del_\varrho -B_{t\, z\phi}\del_z +B_{t\, \phi\varrho}\frac{l_z}{\varrho}  \\
\Gamma^\mathrm{even}_{\mathrm{Sky}, t\, \da\ua}&=
K_t+ B_{t\, \varrho\phi} \del_\varrho +B_{t\, z\phi}\del_z +B_{t\, \phi\varrho}\frac{l_z}{\varrho}  \\
\Gamma^\mathrm{even}_{\mathrm{Sky}, t\, \da\da}&=
-(\del_\varrho M_t) \del_\varrho -(\del_z M_t) \del_z -M_t\bigtriangleup +U_t +\delta_{0t} U_0'-B_{t\, \phi z}\frac{l_z}{\varrho} \\
\Gamma^\mathrm{odd}_{\mathrm{Sky}, t\, \ua\ua}&=
-(\del_\varrho C_{t\, z}) \del_\varrho -(\del_z C_{t\, z} )\del_z -C_{t\, z} \bigtriangleup +\Sigma_{t\, z} +I_{t\, \phi}\frac{l_z}{\varrho} \\
\Gamma^\mathrm{odd}_{\mathrm{Sky}, t\, \ua\da}&=
-(\del_\varrho C_{t\, \varrho}) \del_\varrho -(\del_z C_{t\, \varrho} )\del_z -C_{t\, \varrho} \bigtriangleup +\Sigma_{t\, \varrho} -C_{t\, \varrho} \frac{l_z}{\varrho^2} \\
\Gamma^\mathrm{odd}_{\mathrm{Sky}, t\, \da\ua}&=
-(\del_\varrho C_{t\, \varrho}) \del_\varrho -(\del_z C_{t\, \varrho} )\del_z -C_{t\, \varrho} \bigtriangleup +\Sigma_{t\, \varrho} +C_{t\, \varrho} \frac{l_z}{\varrho^2} \\
\Gamma^\mathrm{odd}_{\mathrm{Sky}, t\, \da\da}&=
(\del_\varrho C_{t\, z}) \del_\varrho +(\del_z C_{t\, z} )\del_z +C_{t\, z} \bigtriangleup -\Sigma_{t\, z} +I_{t\, \phi}\frac{l_z}{\varrho}. 
\end{align}
Here, $l_z = \frac{\del_\phi}{\ii}$ and we defined 
\begin{align}
K_t&\coloneqq\frac{1}{2}\left[(\del_\varrho B_{t\, \varrho\phi}) + (\del_z B_{t\, z\phi})+\frac{1}{\varrho}(B_{t\, \varrho\phi}+B_{t\, \phi\varrho})\right] \notag \\
&= C_t^J \left[ (\del_\varrho J_{t\, \varrho\phi})+(\del_z J_{t\, z\phi}) +\frac{1}{\varrho}(J_{t\, \varrho\phi}+J_{t\, \phi\varrho}) \right].
\end{align}

\subsection{Mean-field potentials in the pairing EDF for axially-symmetric nuclei}
For the pairing energy, we adopt the following density functional:
\begin{equation}
\mathcal{E}_{\rm{pair}}=\dfrac{V_0}{4}\int d\boldsymbol{r} \left\{1-V_1\left[\dfrac{\rho_0(\boldsymbol{r})}{\rho_{\rm{c}}}\right]^\gamma \right\}
\sum_{q=\mathrm{p}, \mathrm{n}}
|\tilde{\rho}^q(\boldsymbol{r})|^2, \label{E_pair}
\end{equation}
where $\tilde{\rho}^q$ is the pairing density~\cite{dob84} and $\rho_{\rm{c}}$ is the saturation density. The contributions to the particle-hole potentials are given by
\begin{equation}
\Gamma^\mathrm{even}_{\mathrm{pair}, t\, \sigma\sigma'}(\boldsymbol{r}) = \delta_{0t} \delta_{\sigma\sigma'} U_{\mathrm{pair}}(\boldsymbol{r}), 
\qquad \Gamma^\mathrm{odd}_{\mathrm{pair}, t\, \sigma\sigma'}(\boldsymbol{r}) = 0,
\end{equation}
where
\begin{equation}
U_{\mathrm{pair}} = -\frac{\gamma V_0 V_1}{4 \rho_{\rm{c}}}\left(\frac{\rho_0}{\rho_{\rm{c}}}\right)^{\gamma-1} \sum_{q={\rm p},{\rm n}} |\tilde{\rho}^q|^2,
\end{equation}
and the particle-particle potential, or the pair Hamiltonian, is given by
\begin{equation}
\tilde{h}^q_{\sigma\sigma'}(\boldsymbol{r})=\delta_{\sigma\sigma'} \tilde{U}^q(\boldsymbol{r}),
\end{equation}
where
\begin{equation}
\tilde{U}^q = \frac{V_0}{2} \left[1-V_1\left(\frac{\rho_0}{\rho_c}\right)^\gamma \right] \tilde{\rho}^q.
\end{equation}
With the axial symmetry, the pairing density in the cylindrical coordinates $\boldsymbol{r}=(\varrho, \phi, z)$ is given by employing the ans\"atze (\ref{wf_axial}) as
\begin{equation}
\tilde{\rho}^q = - \sum_i \left[\varphi_{2,i}^{q+}\varphi_{1,i}^{q+}+\varphi_{2,i}^{q-}\varphi_{1,i}^{q-} \right].
\end{equation}
In the end, one obtains
\begin{equation}
\Gamma^\mathrm{even}_{\mathrm{pair}, t}(\boldsymbol{r})= \delta_{0t}
\begin{bmatrix} 
U_{\mathrm{pair}}(\varrho, z)& 0 \\ 
0 & U_{\mathrm{pair}}(\varrho, z)
\end{bmatrix}, \qquad
\Gamma^\mathrm{odd}_{\mathrm{pair}, t}(\boldsymbol{r})=0,
\end{equation}
and
\begin{equation}
\tilde{h}^q(\boldsymbol{r})=
\begin{bmatrix} 
\tilde{U}^q(\varrho, z)& 0 \\ 
0 & \tilde{U}^q(\varrho, z)
\end{bmatrix}.
\end{equation}

\bibliographystyle{ptephy}
\bibliography{oddHFB_ref}

\end{document}